\documentclass[sn-mathphys,Numbered]{sn-jnl}

\usepackage{graphicx}
\usepackage{svg}
\usepackage{multirow}
\usepackage{amsmath,amssymb,amsfonts}
\usepackage{amsthm}
\usepackage{mathrsfs}
\usepackage[title]{appendix}
\usepackage{xcolor}
\usepackage{textcomp}
\usepackage{manyfoot}
\usepackage{booktabs}
\usepackage{algorithm}
\usepackage{algorithmicx}
\usepackage{algpseudocode}
\usepackage{listings}
\usepackage{comment}

\usepackage{tikz}
\usepackage{pgfplots}
\usepackage{colortbl}
\pgfplotsset{compat=1.17}
\usepackage{multirow}
\usepackage{csquotes}
\MakeOuterQuote{"}
\usepackage{cleveref}

\usepackage{caption}
\usepackage{threeparttable}

\usepackage{soul}

\usepackage{enumitem}

\usepackage{tipx}
\usepackage{adjustbox}
\usepackage{tcolorbox}
\usepackage{listings}
\usepackage{subcaption}
\usepackage{pdfpages}
\usepackage{lineno}

\usepackage[normalem]{ulem}

\usepackage{xspace}
\newcommand{\easer}{EASE$^R$}
\newcommand{\rpb}{$\textrm{RP}^3_{\beta}$}

\newcommand{\sol}[2]{\textcolor{blue}{{\bf [#1]} #2}}

\newcounter{reqcounter}

\definecolor{myBlue}{RGB}{159, 216, 251}
\definecolor{myOrange}{RGB}{255, 152, 115}
\definecolor{myGreen}{RGB}{106, 168, 79}
\definecolor{myRed}{RGB}{153, 0, 0}

\begin{document}



\title{Evaluating ChatGPT as a Recommender System: A Rigorous Approach}


\author*[1]{\fnm{Dario} \sur{Di Palma}}\email{name.surname@poliba.it}

\author*[1]{\fnm{Giovanni Maria} \sur{Biancofiore}}

\author*[1]{\fnm{Vito Walter} \sur{Anelli}}

\author[1]{\fnm{Fedelucio} \sur{Narducci}}

\author[1]{\fnm{Tommaso} \sur{Di Noia}}

\author[1]{\fnm{Eugenio} \sur{Di Sciascio}}

\affil[1]{\orgdiv{Department of Electrical and Information Engineering} \orgname{Polytechnic University of Bari}, \orgaddress{\street{Via Edoardo Orabona 4}, \city{Bari}, \postcode{70126}, \country{Italy}}}

\abstract{
Large Language Models (LLMs) have recently shown impressive abilities in handling various natural language-related tasks. Among different LLMs, current studies have assessed ChatGPT's superior performance across manifold tasks, especially under the zero/few-shot prompting conditions. Given such successes, the Recommender Systems (RSs) research community have started investigating its potential applications within the recommendation scenario. However, although various methods have been proposed to integrate ChatGPT's capabilities into RSs, current research struggles to comprehensively evaluate such models while considering the peculiarities of generative models. Often, evaluations do not consider hallucinations, duplications, and out-of-the-closed domain recommendations and solely focus on accuracy metrics, neglecting the impact on beyond-accuracy facets. 

To bridge this gap, we propose a robust evaluation pipeline to assess ChatGPT's ability as an RS and post-process ChatGPT recommendations to account for these aspects. Through this pipeline, we investigate ChatGPT-3.5 and ChatGPT-4 performance in the recommendation task under the zero-shot condition employing the role-playing prompt. We analyze the model's functionality in three settings: the Top-N Recommendation, the cold-start recommendation, and the re-ranking of a list of recommendations, and in three domains: movies, music, and books. 

The experiments reveal that ChatGPT exhibits higher accuracy than the baselines on books domain. It also excels in re-ranking and cold-start scenarios while maintaining reasonable beyond-accuracy metrics. Furthermore, we measure the similarity between the ChatGPT recommendations and the other recommenders, providing insights about how ChatGPT could be categorized in the realm of recommender systems. The evaluation pipeline is publicly released for future research\footnote{\url{https://github.com/sisinflab/Recommender-ChatGPT}}.
}


\maketitle


\section{Introduction}\label{sec1:Introduction}

The exponential growth of human-generated data, primarily in the form of unstructured text, has fundamentally reshaped the digital landscape. The World Wide Web has become a sprawling repository of information, reflecting a rich tapestry of human and corporate needs, opinions, and knowledge. However, harnessing the value of this vast resource presents a significant challenge.


This challenge necessitates the application of Natural Language Processing (NLP) techniques. NLP encompasses a suite of computational methodologies~\cite{DBLP:journals/access/AbdullahAAAT23} designed to analyze, understand, and extract meaning from natural language. By leveraging these techniques, researchers and practitioners across various fields, including data science, information technology, marketing, and social research, are empowered to unlock the potential of this vast unstructured data repository.


The ease of information sharing and retrieval has drastically accelerated information dissemination, raising concerns about information overload (e.g.~\cite{DBLP:journals/ijhci/Burke12}), quality, privacy, and security.


Different approaches, from Information Retrieval (IR)~\cite{DBLP:journals/ftir/MitraC18} to Recommender Systems (RSs)~\cite{DBLP:reference/sp/2022rsh}, aim to sift through massive data to find valuable information. However, their effectiveness depends largely on efficiently filtering out irrelevant or low-quality information, ensuring users can find reliable and pertinent information quickly.



IR systems have leveraged sophisticated NLP techniques like semantic analysis and natural language understanding to deliver contextually relevant information beyond simple keyword matching. These advancements allow them to understand user queries more deeply.

Recommender Systems have also become increasingly adept at predicting user preferences and delivering personalized content recommendations. By analyzing past behavior, user interactions, and preferences, RSs curate content that aligns with individual tastes and interests. Integrating NLP techniques in RSs enables them to capture user preferences with greater nuance, leading to highly relevant recommendations.

Furthermore, recent studies~\cite{DBLP:journals/corr/abs-2209-01621, DBLP:conf/cikm/ZhangCA0C18, DBLP:conf/sigir/GaoXB20} highlight the advantages of leveraging interactive systems to enhance the quality 
of the user satisfaction and experience. By interacting through human-like dialogues, these systems are adept at fetching more precise data that are closely tailored to individual preferences.
This pivotal shift towards conversational interfaces explains the rising prominence of digital assistants such as Amazon Alexa, Google Assistant, Microsoft Cortana, and Apple Siri \cite{DBLP:conf/cikm/Maarek18}. Their ability to understand and process natural language queries in real time has revolutionized how users interact with technology, making access to information in different forms more intuitive and efficient.

The rise of advanced Language Models, particularly the emergence of Large Language Models (LLMs) like Generative Pre-trained Transformer 3 (GPT-3)~\cite{DBLP:conf/nips/BrownMRSKDNSSAA20}, significantly amplified the potential for interaction with digital systems~\cite{radford2019language} facilitating the access to a quantity of information never seen before, through natural language requests and queries. 
These models have pioneered a new era of machine comprehension and language generation, enabling them to engage in conversations with unprecedented naturalness and depth. Their capabilities extend far beyond simple query processing, facilitating dynamic and nuanced dialogues that mirror human interactions.
Although GPT-3 demonstrates significant improvement and future potential in generating human language text, it was only with the advent of ChatGPT on November 30, 2022\footnote{\url{https://openai.com/blog/chatgpt/}}, that the world saw, for the first time, an AI model capable of communicating and responding to queries like never before, able to cover a wide range of tasks~\cite{DBLP:conf/nips/BrownMRSKDNSSAA20}. 
As a result, researchers have taken a keen interest in exploring ChatGPT's potential in various applications, tailoring it for diverse and specific tasks. In particular, researchers have shown a notable interest in exploring its potential for recommendation tasks~\cite{DBLP:journals/corr/abs-2303-14524, DBLP:journals/corr/abs-2311-02089, DBLP:journals/corr/abs-2308-14296, DBLP:conf/ecir/HouZLLXMZ24}. Despite the growing interest in integrating ChatGPT with recommender systems to enhance accuracy~\cite{DBLP:conf/recsys/DaiSZYSXS0X23, DBLP:journals/corr/abs-2304-10149}, several studies overlook the importance of beyond-accuracy facets. 
While some initial work delves into fairness~\cite{DBLP:conf/recsys/ZhangBZWF023}, aspects such as like recommendation diversity, the system's ability to suggest novel items, and mitigating popularity bias still need to be explored. 
Furthermore, significant challenges persist in evaluating Large Language Model (LLM)-based recommender systems. These include handling hallucinations (generating fictitious information), avoiding duplications (repeated recommendations within a list), and preventing out-of-domain recommendations (suggestions irrelevant to a specific catalogue or evaluation setup), remain largely unaddressed.

These systems hold the potential to revolutionize how users discover content, fostering a personalized and engaging experience. However, to assess how these powerful models understand our needs and preferences, a comprehensive evaluation pipeline needs to step in, acting as a decoder ring to the "black box" of LLM recommendations.
The pipeline should thoroughly analyze the system's performance, objectively evaluating its strengths and areas for improvement.
This multifaceted analysis goes beyond mere accuracy. The ideal pipeline investigates the quality and relevance of recommendations. Not only should it check if the model can avoid the pitfall of repetitive suggestions, but it should steer clear of nonsensical or "hallucinated" content. Ultimately, the goal is to ensure recommendations are accurate, diverse, and truly resonate with the user's unique interests.
The full potential of LLM-powered recommendation systems could be unlocked by constructing such a robust evaluation pipeline.

To bridge these gaps, we present an evaluation pipeline designed to assess ChatGPT's strengths and weaknesses as a recommendation system. 
Employing this pipeline, we delve into the performance of ChatGPT-3.5 and ChatGPT-4 in the recommendation task under the zero-shot role-playing prompt condition. 
This methodology allows for 
evaluating these models' capacity to serve as a standalone Recommender System across three distinct domains: Books, Movies, and Music (i.e. Facebook Books~\cite{DBLP:conf/recsys/MancinoFBMNS23}, MovieLens~\cite{DBLP:journals/tiis/HarperK16}, Last.FM~\cite{DBLP:conf/recsys/CantadorBK11}), and three scenarios: Top-N recommendation, re-ranking a list of recommendations, and cold-start recommendation scenarios.

Additionally, since ChatGPT is a Large Language Model and not a specialized Recommender System, we explore its capability to suggest items by assessing the similarity between its generated recommendation lists and those produced by content-based, collaborative filtering, and hybrid recommenders. This analysis aims to estimate ChatGPT's inclination towards content-based or collaborative filtering methods for generating recommendations. The goal is to gain insights into the underlying approach that ChatGPT mimic in the recommendation process.\\

\noindent Our contributions can be summarized as follows:
\begin{itemize}[leftmargin=7mm]
    \item We define and publicly share with the research community an evaluation pipeline\footnote{\url{https://github.com/sisinflab/Recommender-ChatGPT}} to assess the performance of OpenAI's ChatGPT on recommendation tasks, going beyond mere accuracy metrics to address challenges such as diversity, novelty, popularity bias, and the avoidance of hallucinated content. 
    \item We evaluate the performance of ChatGPT-3.5 and ChatGPT-4 on three distinct recommendation scenarios: classical recommendation, re-ranking a list of recommendations, and cold-start recommendation across three domains: Books, Movies, and Music.
    \item We investigate the underlying methodology employed by ChatGPT in generating recommendations, aiming to understand whether the model demonstrates 
    an inclination towards content-based, collaborative, or hybrid recommenders to gain insights into its recommendation process.
    \newline
\end{itemize}

\noindent The remainder of the paper is structured as follows: Section~\ref{sec2:Background} provides the necessary background, focusing on large language models and the objectives of recommender systems. Section~\ref{sec3:Methods} presents the datasets, briefly introduces prompt engineering, and discusses the choice of prompts and the evaluation pipeline used to obtain the results. Section~\ref{sec4:Experiments} discusses the obtained results and compares ChatGPT's performance with other recommender systems, attempting to identify general patterns. Sections~\ref{sec5:Related} and \ref{sec6:Limitations} present related work on using ChatGPT and other LLMs as recommender systems and discuss the limitations of these approaches. Finally, Section~\ref{sec7:Conclusions} summarizes the results and provides an outlook on future directions.

\section{Background}\label{sec2:Background}
A fertile ground for innovation lies at the intersection of large language models and recommender systems. This section provides a foundation by introducing the background knowledge and core concepts underpinning these two research fields.

\subsection{Large Language Models}\label{sec:background_LLM}
Large Language Models (LLMs) are a cutting-edge type of neural network specifically designed for natural language processing (NLP) tasks. Unlike traditional language models, LLMs leverage the Transformer architecture, empowered by attention mechanisms~\cite{vaswani2017attention}. This allows them to excel at capturing long-range dependencies and sequential patterns within text data~\cite{DBLP:conf/pldi/Xu0NH22}. The defining characteristic of LLMs lies in their massive scale. They boast hundreds of millions, or even billions, of parameters, trained on colossal datasets of text information gathered from the internet and other sources~\cite{zhao2023survey, DBLP:conf/nips/BrownMRSKDNSSAA20}.


At their core, LLMs are probabilistic models designed to estimate the probability distribution of sequences of linguistic units such as words or sentences~\cite{DBLP:journals/pieee/Rosenfeld00}. Within the realm of LLMs based on the Transformer architecture, three primary types emerge: the \textit{Encoder-Decoder}, \textit{Encoder-Only}, and \textit{Decoder-Only} models.

The Encoder-Decoder models, also referred to as sequence-to-sequence models, comprise two key components: an encoder that processes the input sequence, producing a contextualized representation, and a decoder that generates the output sequence based on the encoded input representation~\cite{DBLP:journals/jmlr/RaffelSRLNMZLL20}.
The Encoder-Only models, such as BERT~\cite{DBLP:conf/naacl/DevlinCLT19}, consist solely of an encoder component, which processes the input sequence and generates a contextualized representation for each token. Finally, the Decoder-Only models, also known as causal language models or autoregressive models, consist solely of a decoder component, which generates output sequences token by token based on a learned language and the previously generated tokens~\cite{radford2018improving}.

Among these architectures, ChatGPT employs a decoder-only, autoregressive model. Although it has become popular nowadays, its roots are traced back to the introduction of the first Generative Pre-Trained Transformer (GPT)~\cite{radford2018improving} and GPT-2~\cite{radford2019language}. The GPT models demonstrated how scaling up the number of parameters and the training data size led to remarkable capabilities in language generation that went far beyond traditional language modeling capabilities. Specifically, they exhibited exceptional learning capacity and abilities to perform various natural language tasks. Moreover, the introduction of GPT-3~\cite{brown2020language}, with further increases in model size and training data, showcased how these models learned from few-shot prompts, i.e., a small number of examples provided as part of the input to help understand the task, and perform novel tasks for which they were not explicitly trained.

However, the current state of ChatGPT results from a long history of progress on the pre-training architecture of GPT and the introduction of the fine-tuning approach through Reinforcement Learning from Human Feedback (RLHF)~\cite{DBLP:conf/nips/Ouyang0JAWMZASR22}. This approach first learns broad, general knowledge during the pre-training phase and then adapts this knowledge to specific downstream tasks, i.e., open-ended conversation, through dedicated fine-tuning, enabling generalization and specialization. Currently, OpenAI released a version of ChatGPT based on GPT-3.5~\cite{DBLP:conf/nips/BrownMRSKDNSSAA20} and an advanced one called GTP-4~\cite{achiam2023gpt}. 

By referring to the work of Yang et al.\cite{yang2019xlnet}, which further extends the formalisms introduced by Radford et al.\cite{radford2018improving}, the task of ChatGPT can be summarized as:

\[
\max_\theta \log p_\theta(x_t|\textbf{x})=\sum_{t=1}^T\log p_\theta(x_t|\textbf{x}_{<t})
\]

where $\textbf{x} = [x_1,...,x_T]$ is a textual sequence, i.e., the context, $x_t$ is the next token to predict.

To summarize, ChatGPT, an autoregressive generative model, excels at creating sequential responses like text or code. During training, it learns to model the conditional probability distribution of a word appearing next in a sequence, given the previous words $\textbf{x}$ (like a question or conversation history). At each step, it predicts the most likely word from its vast vocabulary. This process unfolds one word at a time, with each new word $x_{t}$ depending on both the original input and the previously generated sequence $\textbf{x}_{<t}$. It continues until an "end-of-sequence" token marks the completion of the response.

\subsection{Recommender Systems}\label{sec:background_RSs}
Recommender Systems (RSs) are powerful tools that help users navigate vast collections of choices by suggesting items they might be interested in~\cite{zhang2019deep}. They personalize the user experience by learning about user preferences.
This can be done explicitly, where users directly rate or provide feedback on items, or implicitly, by analyzing user behavior like browsing history and purchases~\cite {DBLP:journals/ijhci/Burke12}. 
By building user profiles that capture these preferences, RSs can assess how relevant each item is to a particular user and filter out irrelevant ones.


The recommendation task comes in two main flavors. The first is the rating prediction problem, where the system tries to guess how a user would rate items they have not interacted with yet. The second is the top-\textit{N} recommendation problem, where the goal is to identify a list of $N$ items most relevant to the user~\cite{DBLP:conf/recsys/CremonesiKT10}. These recommended items are typically ranked based on predicted user interest, with the most relevant items at the top.


Both approaches aim to address information overload and suggest relevant items to users. RSs achieve this by modeling a utility maximization problem (e.g., user items' ratings) with the utility function essentially capturing how much a user would enjoy an item. Following the notation introduced by~\citet{DBLP:journals/tkde/AdomaviciusT05}, it considers a set of users ($U$) and all possible recommendable items ($I$). The utility function, denoted as $r: U \times I \rightarrow R$, assigns a numerical value within a specific range to each user-item pair $r(u, i)$, indicating how useful item $i$ would be for user $u$. Formally, the recommendation for user u becomes the item i that maximizes this utility function:
\[
\forall u \in U,\ i^*_u = \operatorname*{arg\,max}_{i \in I} r(u, i).
\]

For each user, an ordered list of $N$ items results from the previous utility maximization problem solving the Top-N Recommendation task. This obtained catalog is commonly known as a ranked items list. 

Depending on the information type employed to compute the utility score of items, the RS can be categorized into three classes: \textit{collaborative filtering}, \textit{content-based filtering}, and \textit{hybrid methods}. Collaborative Filtering (CF) methods produce recommendations based on the underlying principle that users with similar preferences or behavioral patterns in the past will likely have similar preferences for future items~\cite{koren2021advances}. Conversely, Content-based Filtering (CBF) recommenders employ item descriptions (from textual to numerical features) to identify items of particular interest to the user~\cite{pazzani2007content}, suggesting items more similar in content to the ones the user has already liked. Hybrid RSs are designed when both content and collaborative information are jointly exploited to build the list of item recommendations.

\section{Proposed Evaluation Pipeline for ChatGPT for Recommendation}\label{sec3:Methods}
The following sections discuss the methodology and evaluation techniques used in our research. Specifically, Section~\ref{sec:prompt_design} outlines the design and implementation of the used Prompt, detailing the latest prompt engineering strategies and the decision to use Role-Play Prompting. Section~\ref{sec:dataset_pre_processing} examines the datasets employed in the experiments and the necessary pre-processing steps to respect the constraints given by ChatGPT. Section~\ref{sec:eval_pipeline} presents the evaluation pipeline, laying the foundation for the study. Finally, Section~\ref{sec:baseline} and~\ref{sec:metrics} discusses the baselines and metrics for assessing ChatGPT's performance in Recommendation tasks, providing a comprehensive overview of the evaluative criteria.

\subsection{Prompt Design}\label{sec:prompt_design}

The development of Large Language Models (LLMs), especially the GPT series such as GPT-3~\cite{DBLP:conf/nips/BrownMRSKDNSSAA20}, has significantly transformed fields such as technology, education, and content creation. At the core of interactions with these models lies the concept of a "prompt". A prompt is essentially a piece of natural language text given to the model by the user, serving as the initial input or question. This input initiates the model's response process and guides its direction and scope, acting as a catalyst for the output's relevance and utility.


Understanding prompt design is crucial because LLMs learn from extensive datasets, acquiring the ability to understand language and contextual nuances. Furthermore, they demonstrate the ability to perform various tasks by following prompts. Consequently, well-designed prompts enable users to effectively guide the model's output generation process, increasing the likelihood of obtaining relevant and valuable results for the intended task.

However, it was only after the emergence of GPT-3 in 2020~\cite{DBLP:conf/nips/BrownMRSKDNSSAA20} that the significance of crafting effective prompts became evident, giving rise to "prompt engineering" as a dedicated field of study. A field that explores how to prompt LLMs to adapt to new tasks using minimal examples, thereby optimizing the interaction process.

Among the several prompt engineering techniques that can be employed~\cite{DBLP:journals/corr/abs-2402-07927}, our exploration focused on the following:

\textbf{Zero-shot prompting}~\cite{radford2019language}\textbf{.} The most basic and widely used approach. In zero-shot prompting, the context information and task description are provided directly to the LLM without reference examples. This approach leverages the LLM's pre-trained knowledge and capabilities to generate relevant responses based solely on the provided context and instructions.
    
    \adjustbox{center,margin=0.2cm,scale=0.9}{\begin{tcolorbox}[colback=myOrange!5!white, colframe=myOrange, title=Zero-shot prompt]
    I like \{history of the user\}, provide me 50 recommendations.
    \end{tcolorbox}}

\textbf{Few-shot prompting}~\cite{DBLP:conf/nips/BrownMRSKDNSSAA20}\textbf{.} In contrast to zero-shot prompting, few-shot prompting provides a few demonstrations or examples in the prompt to help LLM better understand the user's intent. 
    Therefore, selecting relevant examples plays a vital role in guiding the LLM toward generating personalized and appropriate recommendations.
    
    \adjustbox{center,margin=0.2cm,scale=0.9}{\begin{tcolorbox}[colback=myOrange!5!white, colframe=myOrange, title=Few-shot prompt]
    User's Watched Movie: "The Shawshank Redemption"
    User's Rating: 5 (out of 5)
    Recommended Movie: "The Green Mile"
    Explanation: Both movies are critically acclaimed drama films with themes of hope and resilience in difficult circumstances. Since the user highly rated "The Shawshank Redemption", they might also enjoy "The Green Mile".\\

    User's Watched Movie: "Inception"
    User's Rating: 4
    Recommended Movie: "Interstellar"
    Explanation: Both movies are science-fiction films directed by Christopher Nolan, known for their mind-bending plots and impressive visuals. A user who enjoyed "Inception" is likely to appreciate "Interstellar" as well.\\

    User's Watched Movie: "The Dark Knight"
    User's Rating: 5
    Recommended Movie: 
    \end{tcolorbox}}

\textbf{Chain-of-Thought prompting}~\cite{DBLP:conf/nips/Wei0SBIXCLZ22}\textbf{.} Also known as CoT prompting, this technique is a widely adopted approach for reasoning tasks such as question answering and mathematical inference. This method prompts the LLM to solve problems by breaking them into explicit steps, effectively eliciting the model's ability to reason step-by-step.

    \adjustbox{center,margin=0.2cm,scale=0.9}{\begin{tcolorbox}[colback=myOrange!5!white, colframe=myOrange, title=Chain-of-Thought prompt]
    Instruction: Recommend a book based on the user's preferences.\\
    
    User's Preferences: The user enjoys science fiction, particularly books with strong character development and intricate world-building. They have previously enjoyed "Dune" by Frank Herbert and "Ender's Game" by Orson Scott Card.\\
    
    Chain of Thought: To recommend a suitable book, I need to consider the user's preferences for science fiction, strong character development, and intricate world-building. The user enjoyed "Dune" and "Ender's Game," which suggests they appreciate complex plots and detailed settings. Based on these criteria, I will identify a book that fits within the science fiction genre and has a reputation for similar qualities.\\
    
    1. The user likes science fiction, so I will focus on books within this genre.\\
    2. The user values strong character development, so I will consider books known for their well-developed characters.\\
    3. The user appreciates intricate world-building, so I will look for books with detailed and immersive settings.\\
    4. Based on their enjoyment of "Dune" and "Ender's Game," I'll look for books with complex plots and critical acclaim.\\
    
    Recommendation: "The Left Hand of Darkness" by Ursula K. Le Guin\\
    
    Explanation: "The Left Hand of Darkness" is a renowned science fiction novel known for its detailed world-building and exploration of complex social and cultural issues. Ursula K. Le Guin's character development is exceptional, and the story's setting on the planet Gethen provides a rich and immersive experience. This book should align well with the user's preferences and previous enjoyment of "Dune" and "Ender's Game."
    \end{tcolorbox}}

    
    
\textbf{Role-Playing prompting}~\cite{DBLP:conf/nips/JinC0YF00W23}\textbf{.} Role-playing prompts involve engaging with the language model in a "role-playing" scenario. This approach involves providing auxiliary information in the prompt to describe the role and responsibilities of the role in detail.
    
    \adjustbox{center,margin=0.2cm,scale=0.9}{\begin{tcolorbox}[colback=myOrange!5!white, colframe=myOrange, title=Role-Playing prompt]
    I want you to take on the role of a recommender system. Your goal is to provide tailored book suggestions that users will likely to enjoy based on their history.
    \end{tcolorbox}}

In our study, we aimed to test how well ChatGPT could generate 50 recommendations by looking at a user's history. Initially, our approach employed Few-Shot and Chain-of-Thought prompting techniques to sample ChatGPT's capabilities. However, we encountered a significant challenge due to the constraints on context length imposed by ChatGPT, specifically its version 3.5. Moreover, the vast nature of user histories made the application of few-shot and CoT prompting impractical for our datasets, as these methods often require the inclusion of multiple examples or detailed step-by-step reasoning.

Faced with these limitations, our strategy pivoted towards crafting a straightforward yet impactful zero-shot role-playing prompt that could accommodate our requirement to generate a comprehensive list of 50 recommendations while navigating the constraints of the model's capacity to process extensive information. The refined prompt we developed was as follows:

\adjustbox{center,margin=0.2cm,scale=0.9}{
\begin{tcolorbox}[colback=myBlue!5!white, colframe=myBlue!92!black, title=Top-N Recommendation Prompt]
Given a user, as a Recommender System, please provide only the names of the top 50 recommendations.
You know that the user likes the following items: \{history of the user\}
\end{tcolorbox}}\label{prompt:top_n_rec}

Through heuristic testing (i.e., employing an empirical, trial-and-error approach based on experience and intuition), this prompt emerged as the most effective in eliciting 50 relevant recommendations from ChatGPT. Concurrently, there was a growing body of research examining the efficacy of various prompting strategies. Notably, these studies underscored the effectiveness of Role-Playing Prompting~\cite{DBLP:journals/corr/abs-2401-04997, DBLP:journals/corr/abs-2308-07702}. In our case, we requested the language model to assume a specific role: a Recommender System. This minimalistic yet focused approach to prompting ChatGPT enhanced the precision and relevance of the recommendations generated. 
Encouraged by these findings, we employed Role-Playing Prompting to generate the collections of user recommendations for our study.

\subsection{Pre-processing Phase}\label{sec:dataset_pre_processing}
In the field of Recommender Systems (RSs), the most prevalent approach to evaluate the performance of a model is through offline experiments using pre-existing datasets, often historical data or logs. Among the various datasets, we employed three widely used and publicly available ones spanning different domains, i.e., books, music, and movies. 

Specifically, we used the Facebook Books~\cite{DBLP:conf/recsys/MancinoFBMNS23}, Last.FM~\cite{DBLP:conf/recsys/CantadorBK11}, and MovieLens~\cite{DBLP:journals/tiis/HarperK16} datasets. Although these datasets are well-known in the literature, we had to account for the token limits of ChatGPT API. Due to this limitation, we employed specific preprocessing steps to adapt the user interaction histories to the context length imposed by the API.

Consequently, we applying an iterative-10-core strategy on the users and items within the datasets to keep only users and items with at least ten occurrences. However, after analyzing each dataset, we discovered that for MovieLens, the interaction history of some users exceeded the maximum context length, and further preprocessing was required to reduce the number of interactions. For this reason, we set a threshold of 200 interactions to deal with the context constraints, leading to a modified MovieLens dataset.

The statistics about these datasets before and after pre-processing are reported in Table~\ref{tab:dataset_statistics}.
~

\begin{table*}[t!]
\centering
\caption{Dataset statistics before and after pre-processing with max context length filtering and \scalebox{0.9}{$k-core \geq\ 10$}.}
\label{tab:dataset_statistics}

\resizebox{\textwidth}{!}{

\begin{tabular}{lcccccccccc}

\toprule

\multirow{3}{*}{\centering{\textbf{Dataset}}} & \textbf{Interaction} & \textbf{Users} & \textbf{Items} & \textbf{Sparsity} & \textbf{Interaction} & \textbf{Users} & \textbf{Items} & \textbf{Sparsity} & \multicolumn{2}{c}{\textbf{Content}} \\

\cmidrule(lr){2-5} 
\cmidrule(lr){6-9}
\cmidrule(lr){10-11}

& \multicolumn{4}{c}{\textit{before pre-processing}} & \multicolumn{4}{c}{\textit{after pre-processing}} & \textit{\small{type}} & \textit{\small{features}}\\
\cmidrule(lr){2-5}
\cmidrule(lr){6-9}
\cmidrule(lr){10-11}

\textbf{MovieLens} & 100836 & 610 & 9724 & 98.30\% & 42456 & 603 & 1862 & 96.22\% & genre & 20 \\
\textbf{Last.FM} & 86608 & 1892 & 12133 & 99.62\% & 49171 & 1797 & 1507 & 98.18\% & genre & 9748 \\
\textbf{FB Books} & 18978 & 1398 & 2234 & 99.74\% & 13117 & 1398 & 2234 & 99.58\% & \small{genre, author} & 1970  \\
\bottomrule
\end{tabular}
}
\end{table*}

A concise overview of each dataset is provided below:\\
~\textbf{Facebook Books Dataset.} Facebook Books\footnote{\url{https://github.com/sisinflab/LinkedDatasets/}} is a dataset released for the Linked Open Data Challenge 2015\footnote{\url{https://2015.eswc-conferences.org/program/semwebeval.html}} covering the book domain. It contains implicit feedback and item-feature mappings to DBpedia for each book, enabling the retrieval of data content such as book genres and relevant author information.

\textbf{Last.FM Dataset.} The Last.FM dataset corresponds to user-artist plays on the Last.FM online music system released during the HETRec2011 Workshop\footnote{\url{https://grouplens.org/datasets/hetrec-2011/}}~\cite{DBLP:conf/recsys/CantadorBK11}. It contains information on social networking, tagging, artists, and music listening from 2000 users.

\textbf{MovieLens Dataset.} The MovieLens dataset is widely exploited in the Recommender Systems community~\cite{DBLP:journals/tiis/HarperK16}. Different versions are available\footnote{\url{https://grouplens.org/datasets/movielens/}}, but the one used in our study was collected from the MovieLens 100k and contains ratings for movies on a 1-5 scale.

\subsection{Evaluation Pipeline}\label{sec:eval_pipeline}
As highlighted in Section~\ref{sec:dataset_pre_processing}, the most prevalent approach to evaluate the performance of recommender systems is through offline experiments using pre-existing datasets. Specifically, the models are trained on these datasets, learning to represent items by their unique itemIDs. During inference, the same itemIDs are used for prediction. Consequently, the classical evaluation pipeline for recommender systems utilizes the list of predicted recommendations based on itemIDs to assess the models' performance.

However, Large Language Models (LLMs) do not generate item IDs; instead, they produce text representing the item names. Consequently, evaluating the performance of ChatGPT in generating recommendations poses a significant challenge due to the complex task of matching the item names provided by the model with those present in the datasets, especially when dealing with noisy datasets where item names are not accurately represented.

\begin{table*}[t!]
\centering
\caption{\small{Comparative assessment of ChatGPT-3.5: \textbf{Old vs. New evaluation pipeline} across four principal metrics: Accuracy, Diversity, Novelty, and Bias. The direction of the arrows (\textuparrow for higher, \textdownarrow for lower) denotes the preferred values for each metric. The best values are in bold. The second-best values are underlined.}}
\label{tab:new_vs_old_pipeline}

\resizebox{\textwidth}{!}{ 

\begin{tabular}{clccccccccc}

\toprule

& \textbf{Evaluation} & \textbf{nDCG} \textupfullarrow & \textbf{Recall} \textupfullarrow & \textbf{Precision} \textupfullarrow & \textbf{Gini} \textupfullarrow & \textbf{ItemCV} \textupfullarrow & \textbf{EPC} \textupfullarrow & \textbf{EFD} \textupfullarrow & \textbf{APLT} \textupfullarrow & \textbf{ARP} \textdownfullarrow \\

\midrule

\multirow{2}{*}{\adjustbox{angle=90, scale=0.7}{\textbf{Books}}}

& New Pipeline & \textbf{0.0615} & \textbf{0.0917} & \textbf{0.0194} & 0.0389 & 616 & \textbf{0.0225} & \textbf{0.2176} & \textbf{0.1185} & 63.0937 \\

& Old Pipeline & 0.0574 & 0.0863 & 0.0184 & \textbf{0.0439} & \textbf{689} & 0.0210 & 0.2032 & 0.1737 & \textbf{58.1202} \\

\midrule

\multirow{2}{*}{\adjustbox{angle=90, scale=0.7}{\textbf{Music}}} 

& New Pipeline & 0.2253 & 0.1835 & 0.1299    & 0.1495 & 827          & 0.1532 & 1.4855 & \textbf{0.1078} & 106.9603 \\

& Old Pipeline & \textbf{0.2268} & \textbf{0.1853} & \textbf{0.1311}    & \textbf{0.1517} & \textbf{833}          & \textbf{0.1535} & \textbf{1.4886} & 0.1077 &\textbf{ 106.5101} \\

\midrule

\multirow{2}{*}{\adjustbox{angle=90, scale=0.7}{\textbf{Movies}}}

& New Pipeline & \textbf{0.1519} & \textbf{0.0815} & \textbf{0.1298} & 0.0367 & 406 & \textbf{0.1208} & \textbf{1.3507} & \textbf{0.0435} & \textbf{122.0000} \\

& Old Pipeline & 0.1514 & \textbf{0.0815} & 0.1297 & \textbf{0.0371} & \textbf{408} & 0.1203 & 1.3449 & 0.0427 & 121.6523 \\

\bottomrule

\end{tabular}

} 

\end{table*}

An initial attempt was made to evaluate ChatGPT's performance using a simple pattern-matching approach. Despite the seemingly promising results, the evaluation method was fundamentally flawed, leading to incorrect conclusions due to various reasons. The primary factor contributing to the inaccurate evaluation stemmed from the failure to properly match items within the dataset, see~\ref{appendix:ChatGPT_normal_response_example} for an example of a typical response from ChatGPT. Specifically, (i.) the pattern-matching algorithm failed to account for variations in item representations, such as matching "1Q84" or "The Twelve Monkeys" by ChatGPT with "IQ84" and "12 Monkeys" in the Facebook Books and MovieLens datasets; (ii.) it incorrectly matched items, e.g., associating "The Dark Tower III" or "Zombieland: Double Tap (2019)" by ChatGPT with "The Dark Tower II" and "Zombieland (2009)" in the datasets.

Furthermore, as known in the literature, generative models like ChatGPT tend to hallucinate under certain conditions~\cite{DBLP:conf/cikm/ChenFYWFL0LX23, DBLP:conf/acl/NieYWPL19, DBLP:journals/csur/JiLFYSXIBMF23}. Upon further analysis, we discovered that in our study, ChatGPT would arbitrarily repeat the same item~\ref{appendix:ChatGPT_duplication_error}, or sequence of items~\ref{appendix:ChatGPT_extreme_duplication}, after recommending a certain number of them until reaching the imposed limit of 50 recommendations. This behavior led to misevaluation by either improving or worsening the evaluation metrics, depending on the specific items repeated and the metric studied.

\begin{figure}
    \centering
    \includegraphics[width=\textwidth]{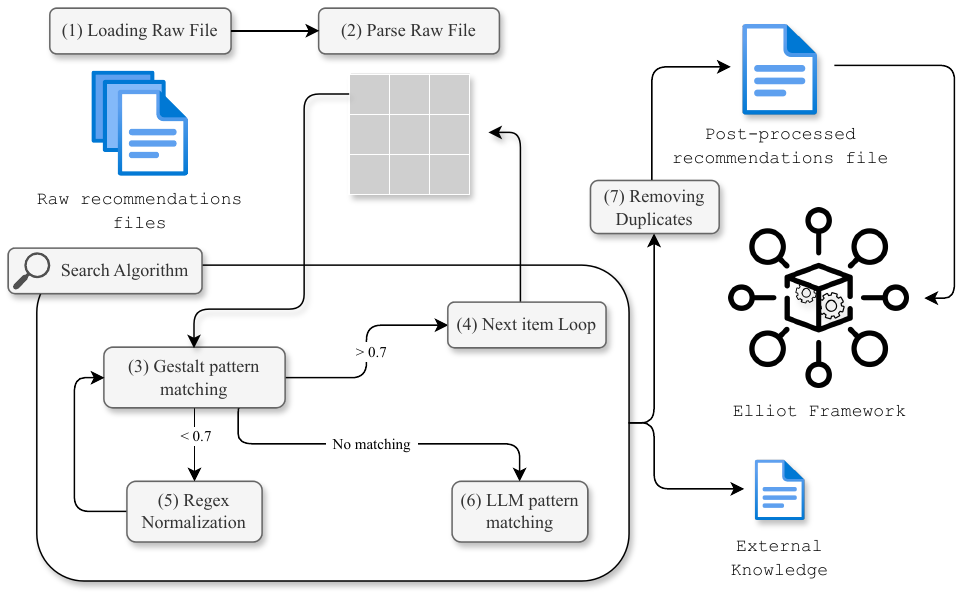}
    \caption{Multi-Stage Evaluation Pipeline for Assessing ChatGPT Recommendation Performance.}
    \label{fig:EvalPipeline}
    \vspace{-1em}
\end{figure}

Motivated by these findings, this work present in Figure~\ref{fig:EvalPipeline} a comprehensive pipeline to evaluate ChatGPT's recommendation capabilities while accounting the specific characteristics of closed-domain datasets, i.e., Facebook Books, MovieLens and Last.FM. 

Specifically, the pipeline employs a multi-stage process that commences by (1) loading a folder containing the collections of the generated text files, each representing a list of 50 recommendations per user under evaluation. For each file and line representing an item, a multi-step search algorithm is utilized to probe for a corresponding item in the dataset.

The algorithm begins with an initial parsing step (2) aimed at separating the names of the recommended items from any additional contextual information within the sentence. Subsequently, a (3) Gestalt pattern matching technique~\cite{DBLP:journals/iandc/GiulianoJKMS61, pythonDifflibHelpers} is used to search for the closest match between the recommended item and the reference dataset. If the similarity score from the pattern matching exceeds a predetermined threshold of 70\%, i.e., empirical value capable of covering many items across all three datasets, the search algorithm proceeds to the next item (4) and repeat the loop. Otherwise, a regular expression (5) is applied to normalize the string, removing accents, converting characters to lowercase, eliminating punctuation, and trimming extra whitespace. This step is applied only in the second case because, in some instances, these elements for certain items are essential to reaching the threshold, and only for a narrow set of items does this step aid in finding a match.

Therefore, for the remaining unrecognized items, a final step leverages a quantized 7-billion parameter LLM (6), i.e., mistral-7b~\cite{jiang2023mistral, huggingfaceTheBlokeMistral7BInstructv02AWQHugging}, which, through few-shot prompt engineering techniques, refer to~\ref{appendix:Mistral_few_shot}, identifies items that are impossible to recognize using classical methods, e.g., The "Twelve Monkeys" with "12 Monkeys" or "The Secret of My Success" with "Secret of My Succe\$s, The". Only after this last step, if no match is found in the dataset, the item is flagged as "External Knowledge" representing information known to the LLM due to its training but absent from the reference dataset. As a last step (7), the duplicate items are removed for each user, and only users with at least 20 valid recommendations are used for the overall evaluation. Additionally, the algorithm employs a caching strategy to speed up the search process and reduce the reliance on the LLM, which is utilized for only 5.56\% of the total items, thereby minimizing the impact and improving search speed. 

A summary of the impact of this pipeline is presented as follows:
\begin{itemize}
    \item For the Facebook Books dataset, on average, 3\% of users had less than 20 unique recommendations, and approximately 2800 items were labeled as external knowledge or duplicates.
    \item The Last.FM dataset exhibited a mean of 1.11\% of users with fewer than 20 unique recommendations, along with an average of 919 items identified as external knowledge or duplicates.
    \item In the case of the MovieLens dataset, the mean percentage of users with less than 20 unique recommendations stood at 2.66\%, while an average of 252 items were classified as external knowledge or duplicates.
\end{itemize}

From the perspective of the evaluation metrics, the impact of the new evaluation pipeline is shown in Table~\ref{tab:new_vs_old_pipeline}. Although for some metrics and datasets, the difference between the old and new pipeline could appear minimal, the importance of having metrics based on a rigorous post-processing approach cannot be overstated. An aspect that currently needs to be improved in the state-of-the-art evaluation of ChatGPT as an RS.


~\begin{table}[t!]
\centering
\caption{Overview of beyond-accuracy metrics~\cite{DBLP:conf/um/AnelliBNJP22}}
\label{tab:beyond_accuracy_table}

\begin{tabular}{@{}p{0.25\textwidth}p{0.15\textwidth}p{0.55\textwidth}@{}}
\toprule
\textbf{Aspect} & \textbf{Metric} & \textbf{Description} \\
\midrule
\multirow{2}{=}{Diversity} & ItemCV & 
Item Coverage (ItemCV) measures how many items ever appear in the top-$n$ recommendations of users. \\
\cmidrule{2-3}
& Gini & Gini Index: A measure of statistical dispersion intended to represent the inequality of a distribution. The Gini Index ranges between 0 and 1, where a higher value indicates greater concentration of recommendations, e.g., on popular items~\cite{DBLP:journals/umuai/JannachLKJ15}. For interpretation convenience and to associate higher values with less concentrated recommendations, we report \(1-\text{Gini}\). \\
\midrule
\multirow{2}{=}{Novelty} & EFD & Expected Free Discovery (EFD): A novelty measure based on the inverse collection frequency, expressing the algorithm's ability to recommend relevant long-tail items~\cite{DBLP:conf/recsys/VargasC11}. \\
\cmidrule{2-3}
& EPC & Expected Popularity Complement (EPC): This metric measures the "number of unseen items now seen" as a way to quantify novelty in recommendations~\cite{DBLP:conf/recsys/VargasC11}. \\
\midrule
\multirow{2}{=}{Popularity Bias} & APLT & Average Popularity of Long-Tail Items (APLT): Measures the average popularity of long-tail items in the top-$n$ recommendations~\cite{DBLP:conf/flairs/AbdollahpouriBM19}. \\
\cmidrule{2-3}
& ARP & Average Rating-based Popularity (ARP): Computes the popularity of items in a recommendation list based on the number of interactions each item has in the training data~\cite{DBLP:journals/umuai/JannachLKJ15}. \\
\bottomrule
\end{tabular}

\end{table}

\subsection{Assessing ChatGPT performance}\label{sec:metrics}
To ensure a comprehensive evaluation of ChatGPT in a recommendation scenario, we assess its performance from several perspectives. Specifically, we collect accuracy metrics and "beyond-accuracy" metrics commonly used in the literature to assess additional quality aspects of recommendation lists.

\begin{itemize}
    \item In terms of \textit{accuracy metrics}, we are interested in evaluating whether the recommender system predicts items that the user will interact with. To assess this, we use two standard metrics: Precision and Recall~\cite{DBLP:reference/sp/2022rsh, DBLP:books/daglib/0022145}. Higher values of Precision and Recall indicate that the recommender system recommends a larger number of relevant items to the users. However, beyond just considering the relevance of the recommendations, our evaluation also considers the ranking of the items using the Normalized Discounted Cumulative Gain (nDCG)~\cite{DBLP:journals/tois/JarvelinK02}. The nDCG metric measures not only the relevance of the recommended items but also their positions in the ranked list, giving more weight to highly relevant items ranked higher. A higher nDCG value indicates better recommendation lists.
    \item Considering \textit{beyond-accuracy metrics}, we measured a broader range of metrics related to diversity, novelty, and popularity bias in the recommendation lists. The details of the specific metrics considered are provided in Table~\ref{tab:beyond_accuracy_table}.
\end{itemize}

\section{Experiments}\label{sec4:Experiments}
This section discusses the experimental setting and the results obtained from evaluating ChatGPT as a Recommender System (RS). Each experiment is designed to address the following research questions: 

\begin{enumerate}[label=\textbf{RQ\arabic*}., align=left, leftmargin=*]\label{enum:RQs}
    \item How do the ChatGPT models perform in Top-N Recommendation tasks compared to traditional recommender systems? (Discussed in Section~\ref{sec:classical_rec_performance})
    \item Can ChatGPT effectively re-rank and improve recommendations by leveraging user history? (Explored in Section~\ref{sec:reranking_evaluation})
    \item How does ChatGPT perform in cold-start scenarios compared to the Top-N recommender systems? (Investigated in Section~\ref{sec:cold_start_scanarios})
    \item To what extent do the recommendation lists generated by ChatGPT align with those produced by Collaborative Filtering and Content-based Recommender Systems? (Examined in Section~\ref{sec:rec_similarities})
\end{enumerate}

\subsection{Experimental Setting}\label{sec:exp_settings}
The experiments compare the performance of ChatGPT-3.5 and ChatGPT-4 models against the recommender systems presented in Section~\ref{sec:baseline}. To ensure a fair comparison, we perform a comprehensive Bayesian hyper-parameter optimization to find the optimal configuration for each baseline based on each training data. Furthermore, for complete reproducibility, we utilize the Elliot recommendation framework~\cite{DBLP:conf/sigir/AnelliBFMMPDN21}, making code, data, and instructions to replicate the experiments publicly available\footnote{\url{https://github.com/sisinflab/Recommender-ChatGPT}}.

To evaluate the models, we employ the all unrated items protocol~\cite{DBLP:conf/recsys/Steck13}, where, for each user, the set of recommendable items comprises all items except the ones the user has previously interacted with. The datasets are split into training and test sets, with 80\% of the user-item interactions used for training and the remaining 20\% held out for testing. 

While recommendation baselines can be implemented using open-source framework like Elliot, generating recommendations from the ChatGPT models requires utilizing the OpenAI API. Specifically, for each user in each dataset (as described in Section~\ref{sec:prompt_design}), we craft a prompt to obtain a list of recommended items. We sent approximately 11,500 API requests for each language model, with ChatGPT-3.5 taking over six days to complete, while ChatGPT-4 took roughly eleven days. For the sake of completeness, since the ChatGPT models are closed-source and proprietary, also the overall costs should be reported. Specifically, the combined cost across both models amounts to approximately \$400.

The generated recommended lists (as reported in Section~\ref{sec:eval_pipeline},) are raw text outputs that require robust post-processing steps, and the computational time required for this post-processing step varies significantly depending on the size of the dataset. For the relatively small Facebook Books dataset, post-processing takes approximately 77 hours and 46 minutes, equivalent to three days. The larger MovieLens dataset requires about 100 hours and 50 minutes of processing time, or four days. The most substantial processing demands are for the large and heterogeneous HetRec dataset, which takes 145 hours and 20 minutes, roughly six days.

Finally, we evaluate the performance of the proposed models and the baseline models on the test set. We report the results using various evaluation metrics, considering each user's top 10 recommended items (i.e., cutoff at 10), commonly used as a benchmark for evaluating the effectiveness of RSs~\cite{DBLP:conf/recsys/PetrovM23, DBLP:conf/recsys/MancinoFBMNS23, DBLP:conf/recsys/RendleKZA20, DBLP:conf/www/HeLZNHC17}. For each experimental setting, we provide a detailed explanation and analysis of the results. The exploration is conducted separately for each domain, and a comprehensive conclusion is drawn to address each \textit{Research Question (RQ).}

\subsection{Baseline Models}\label{sec:baseline}
This section presents all the models used in the evaluation to position ChatGPT among the current state-of-the-art recommenders. To ensure the reproducibility of our experiments, we employ the Elliot framework~\cite{DBLP:conf/sigir/AnelliBFMMPDN21} as a backbone for testing the RSs. 
We partition the models into Collaborative Filtering, Content-Based Filtering, and Non-personalized (for a theoretical introduction, refer to Section~\ref{sec:background_RSs}):

\paragraph{Collaborative Filtering}
~\begin{enumerate}[label=(\roman*)]
\item \easer, presented at WWW\textprimstress19~\cite{DBLP:conf/www/Steck19}, is a linear model without hidden layers, which combines the strengths of matrix factorization and neighborhood-based collaborative filtering approaches. It extends the classic matrix factorization approach by incorporating item-item and user-user similarity information, which helps alleviate the data sparsity problem and improve recommendation accuracy.

\item \rpb, introduced in TOIS\textprimstress16~\cite{DBLP:journals/tiis/PaudelCNB17}, is a graph-based recommendation algorithm that employs random walks and a reranking strategy based on the $P^3$ paths~\cite{DBLP:conf/www/CooperLRS14}. It demonstrates improvement in recommendation accuracy and diversity compared to traditional vertex ranking algorithms, such as PageRank and $P^3\alpha$.

\item ItemKNN (Item-based K-Nearest Neighbors)~\cite{DBLP:conf/sigecom/SarwarKKR00} is a memory-based collaborative filtering algorithm that makes predictions based on item similarities. It finds the k-most similar items (neighbors) to a target item using similarity measures such as cosine vector similarity~\cite{DBLP:journals/cacm/BalabanovicS97} or Pearson correlation coefficient~\cite{DBLP:journals/ir/HerlockerKR02}. The preference scores for the target item are then estimated as a weighted sum of the preference scores of the neighboring items, where the similarities between the items determine the weights.

\item UserKNN (User-based K-Nearest Neighbors)~\cite{DBLP:conf/uai/BreeseHK98} is a memory-based collaborative filtering algorithm that makes predictions based on user similarities, akin to how ItemKNN operates on item similarities. It finds the k-most similar users (neighbors) to a target user and estimates the preference score for a target user-item pair as a weighted sum of the neighboring users' preferences for the same item, with weights determined by user similarities.

\item LightGCN, presented at SIGIR\textprimstress20~\cite{DBLP:conf/sigir/0001DWLZ020}, is a graph convolutional network (GCN) based recommender system that simplifies the design of GCNs for recommendation tasks. It employs a light graph convolution operation and a simple layer combination strategy, significantly reducing the model complexity compared to previous graph-based approaches. LightGCN demonstrates several strengths: simplicity, efficient training, better generalization ability, and effectiveness.

\item MF2020~\cite{DBLP:conf/recsys/RendleKZA20}, introduced at RecSys\textprimstress20, is a state-of-the-art advancement in Matrix Factorization. This collaborative filtering technique approximates the user-item interaction matrix as the product of low-rank user and item factor matrices. While MF methods originated from Singular Value Decomposition (SVD) in the late 1990s~\cite{DBLP:conf/icml/BillsusP98}, MF2020 introduces novel enhancements leveraging the latest research developments.

\item  NeuMF~\cite{DBLP:conf/www/HeLZNHC17} (or more broadly NCF), proposed at WWW\textprimstress17, is an influential neural network-based recommendation model that combines matrix factorization and deep learning. It uses a multi-layer perceptron to model non-linear user-item interactions, combining the output with the matrix factorization component. NeuMF has been widely adopted as a strong neural baseline due to its ability to capture complex patterns in user-item interactions.
\end{enumerate}

\paragraph{Content-Based Filtering}
~\begin{enumerate}[label=(\roman*), start=8]
\item VSM (Vector Space Model fed with Knowledge Graph information)~\cite{DBLP:conf/i-semantics/NoiaMORZ12} is a content-based recommender system that represents items as vectors in a high-dimensional space, where each dimension corresponds to a feature derived from a Knowledge Graph. It computes the relevance score between a user and an item as the cosine similarity between the user's profile vector and the item's feature vector. Recommendations are generated by ranking items based on their relevance scores.

\item AttributeItemKNN~\cite{DBLP:conf/recsys/GantnerRFS11} is a content-based recommender representing items as TF-IDF weighted attribute vectors. It calculates item similarities using vector metrics like cosine similarity and generates recommendations based on the k-nearest neighbour items to a user's preference profile or target item.

\item AttributeUserKNN~\cite{DBLP:conf/recsys/GantnerRFS11} is similar to AttributeItemKNN, but instead of items, it represents the users as a vector of weights.
\end{enumerate}

\paragraph{Non-personalized}
~\begin{enumerate}[label=(\roman*),start=11]
    \item Random Models. A non-personalized algorithm that randomly recommends items according to a uniform distribution. It is used to provide a reference point for beyond-accuracy measures.
    \item MostPop Model. A Non-personalized algorithm that recommends the most popular items to each user, where the number of observed interactions in the training data defines popularity. Typically, it shows good performance on statistical biases datatest~\cite{DBLP:journals/ir/BelloginCC17}; for this reason, it is an important baseline to discover popularity biased dataset~\cite{DBLP:conf/recsys/CremonesiKT10}.
\end{enumerate}


\subsection{Top-N Recommendation Performance}\label{sec:classical_rec_performance}
ChatGPT, as emphasized in Section~\ref{sec1:Introduction}, has gained significant attention due to its capability to perform a wide range of tasks, including generating recommendations. This section aims to evaluate the performance of the ChatGPT model on classical Top-N Recommendation tasks and compare it with traditional recommendation systems, thereby addressing the research question: 

\noindent\textit{\ul{RQ1. How do the ChatGPT models perform in Top-N Recommendation tasks compared to traditional recommender systems?}}

To evaluate ChatGPT as a Recommender System, we follow the standard evaluation practice~\cite{DBLP:reference/sp/GunawardanaSY22}, selecting the best model based on accuracy metrics (e.g., nDCG, Recall, or Precision). These metrics measure the quality of the recommended items in terms of their relevance and ranking. Additionally, we have included beyond-accuracy metrics in our study to gain deeper insights into the performance and characteristics of ChatGPT as an RS~\cite{DBLP:conf/recsys/PaparellaPAN23}. These beyond-accuracy metrics include Diversity, Novelty, and Popularity Bias of the recommendations. Refer to the explanations provided in Section~\ref{sec:metrics} for an interpretation of the metrics presented in the following.

Furthermore, a statistical significance test was conducted using the paired t-test \cite{student1908probable}. In the following tables, most values are statistically significant ($p < 0.05$), except for those marked with * for ChatGPT-3.5 and \dag\ for ChatGPT-4. We will use the same notation (* and \dag) to facilitate the discussion to highlight when the metrics are not statistically significant.\\

\begin{table*}[t!]
\centering

\caption{\small{\textbf{Top-N Evaluation} of ChatGPT-3.5 and ChatGPT-4 Performance on Four Key Metrics: Accuracy, Diversity, Novelty, and Popularity Bias on \textbf{Facebook Books}. The direction of the arrows (\textuparrow for higher, \textdownarrow for lower) denotes the preferred values for each metric. The best values are in bold. The second-best values are underlined. The results are ordered by nDCG. The baselines are statistically significant based on paired t-tests ($p < 0.05$) except for the values denoted with *(for ChatGPT-3.5) and \dag(for ChatGPT-4). Best- CF and CBF correspond to the best Collaborative Filtering and Content-Based Filtering based on the nDCG.}}

\label{tab:classicmetrics_book_10}

\resizebox{\textwidth}{!}{ 

\begin{tabular}{lllllclllc}

\toprule
\multicolumn{10}{c}{\textbf{Facebook Books}}\\
\toprule

\textbf{Model} & \textbf{nDCG} \textupfullarrow & \textbf{Recall} \textupfullarrow & \textbf{Precision} \textupfullarrow & \textbf{Gini} \textupfullarrow & \textbf{ItemCV} \textupfullarrow & \textbf{EPC} \textupfullarrow & \textbf{EFD} \textupfullarrow & \textbf{APLT} \textupfullarrow & \textbf{ARP} \textdownfullarrow \\

\midrule

Random      & 0.0019          & 0.0034          & 0.0008          & 0.7753 & 2230 & 0.0007          & 0.0078          & 0.6874 & 5.7186 \\
MostPop          & 0.0091          & 0.0137          & 0.0033          & 0.0045          & 17            & 0.0031          & 0.0228          & 0.0000          & 138.3632        \\


LightGCN         & 0.0105          & 0.0171          & 0.0038          & 0.0053          & 112           & 0.0035          & 0.0269          & 0.0132          & 134.0763        \\
NeuMF            & 0.0167          & 0.0243          & 0.0057          & \ul{0.3336}          & 1563          & 0.0065          & 0.0661          & 0.2444\dag          & 16.0072         \\
\easer            & 0.0188          & 0.0313          & 0.0071          & 0.0111          & 228           & 0.0066          & 0.0547          & 0.0032          & 125.2026        \\
ItemKNN          & 0.0288          & 0.0408          & 0.0086          & \textbf{0.5293}    & \textbf{2141}    & 0.0104          & 0.1099          & \textbf{0.5974}    & 24.9652         \\
MF2020           & 0.0317          & 0.0592          & 0.0133          & 0.0044          & 15            & 0.0116          & 0.0953          & 0.0000          & 114.0167        \\
UserKNN          & 0.0320          & 0.0468          & 0.0098          & 0.1564          & 1372          & 0.0115          & 0.1065          & 0.0852          & 55.2988         \\
\rpb\ \textsuperscript{\textbf{BestCF}}          & 0.0379          & 0.0568          & 0.0120          & 0.3063          & \ul{1888}          & 0.0138          & 0.1357          & 0.3308          & 44.1225*         \\
AttributeUserKNN & 0.0402          & 0.0593          & 0.0133          & 0.0918          & 945           & 0.0152          & 0.1414          & 0.0466          & 64.2887         \\
VSM              & 0.0458          & 0.0785          & 0.0172          & 0.2478          & 1389          & 0.0173          & 0.1913          & 0.5761          & \ul{7.3705}          \\
AttributeItemKNN\ \textsuperscript{\textbf{BestCBF}} & 0.0479          & 0.0705          & 0.0155          & 0.2824          & 1510          & 0.0182          & 0.2019          & \ul{0.5879}          & \textbf{7.1801}    \\

\cmidrule(lr){1-10}

ChatGPT-3.5      & \ul{0.0668}    & \ul{0.0936}    & \ul{0.0205}    & 0.0713          & 853           & \ul{0.0250}    & \ul{0.2480}    & 0.1870          & 46.3236         \\

ChatGPT-4        & \textbf{0.0932} & \textbf{0.1283} & \textbf{0.0283} & 0.1050          & 1004          & \textbf{0.0353} & \textbf{0.3486} & 0.2424          & 40.0319 \\

\bottomrule

\end{tabular}

} 

\end{table*}
\paragraph{Books Domain}
~Table~\ref{tab:classicmetrics_book_10} summarizes the performance of ChatGPT-3.5, ChatGPT-4 models, and all comparison baselines on the Facebook Books dataset, and we have the following observations:

\textbf{Accuracy.} ChatGPT-4 achieves the best performance across all accuracy metrics (i.e., nDCG, Recall, and Precision), with ChatGPT-3.5 being the runner-up. \textit{Quantitatively}, compared to the strongest baseline (AttributeItemKNN), ChatGPT-4 shows improvements of +94.57\% in nDCG, +81.9\% in Recall, and +82.6\% in Precision. Meanwhile, ChatGPT-3.5 shows improvements of +39.5\% in nDCG, +32.8\% in Recall, and +32.3\% in Precision. Furthermore, ChatGPT-4 over ChatGPT-3.5 demonstrates an improvement of +39.5\% in nDCG, +37.1\% in Recall, and +38\% in Precision. These results highlight ChatGPT's capabilities in recommending relevant items and creating personalized lists tailored to each user's preferences, with ChatGPT-4 outperforming ChatGPT-3.5.

\textbf{Diversity.} ChatGPT-4 exhibits a deterioration of -80.2\% in Gini and -53.1\% in ItemCV compared to the strongest baseline, ItemKNN. However, when compared to MostPop (which has an ItemCV of 15 items versus 1004 for ChatGPT-4, and 853 for ChatGPT-3.5), both ChatGPT-4 and ChatGPT-3.5 demonstrate the ability to suggest a diverse range of items beyond just the most popular ones, albeit focusing on a subset of the catalog. This suggests that while ChatGPT may not achieve the same level of diversity as the baseline, it still offers a more varied set of recommendations than simply suggesting the most popular items.

\textbf{Novelty.} Masured by EPC and EFD, ChatGPT-4 emerges as the top-performing model, followed by ChatGPT-3.5. Compared to the best baseline, AttributeItemKNN, ChatGPT-4 demonstrates substantial improvements of +93.96\% in EPC and +72.66\% in EFD. Moreover, it outperforms ChatGPT-3.5 by +41.2\% in EPC and +40.6\% in EFD. These results highlight the ability of the ChatGPT models, particularly ChatGPT-4, to recommend novel and diverse items within a single list, deviating from conventional recommendations and catering to users' unique preferences.

\textbf{Popularity Bias.} Measured by APLT and ARP, both ChatGPT-4 and ChatGPT-3.5 exhibit a deterioration when compared to the strongest baselines. Specifically, ChatGPT-4 shows a -59.4\% decrease in APLT compared to ItemKNN, and a -457.5\% decrease in ARP compared to AttributeItemKNN. A similar degradation is observed for ChatGPT-3.5. 
This decline in performance for APLT and ARP is to be expected, as these metrics are correlated with ItemCV;
models recommending fewer items tend to have a higher popularity bias, recommending the same popular items more frequently. Consequently, while ChatGPT excels in other areas, its tendency to focus on a subset of the catalog leads to a trade-off in terms of popularity bias.\\

\begin{table*}[t!]
\centering

\caption{\small{\textbf{Top-N Evaluation} of ChatGPT-3.5 and ChatGPT-4 Performance on Four Key Metrics: Accuracy, Diversity, Novelty, and Popularity Bias on \textbf{Last.FM}. The direction of the arrows (\textuparrow for higher, \textdownarrow for lower) denotes the preferred values for each metric. The best values are in bold. The second-best values are underlined. The results are ordered by nDCG. The baselines are statistically significant based on paired t-tests ($p < 0.05$) except for the values denoted with *(for ChatGPT-3.5) and \dag(for ChatGPT-4). Best- CF and CBF correspond to the best Collaborative Filtering and Content-Based Filtering based on the nDCG.}}

\label{tab:classicmetrics_music_10}

\resizebox{\textwidth}{!}{ 

\begin{tabular}{lllllclllc}

\toprule
\multicolumn{10}{c}{\textbf{Last.FM}}\\
\toprule

\textbf{Model} & \textbf{nDCG} \textupfullarrow & \textbf{Recall} \textupfullarrow & \textbf{Precision} \textupfullarrow & \textbf{Gini} \textupfullarrow & \textbf{ItemCV} \textupfullarrow & \textbf{EPC} \textupfullarrow & \textbf{EFD} \textupfullarrow & \textbf{APLT} \textupfullarrow & \textbf{ARP} \textdownfullarrow \\

\midrule

Random                                         & 0.0044           & 0.0068          & 0.0052          & 0.8398 & 1507 & 0.0045          & 0.0478          & 0.5678 & 31.6844 \\
NeuMF                                               & 0.1005          & 0.1133          & 0.0860          & \textbf{0.5049}    & \textbf{1492}    & 0.0804          & 0.7848          & \ul{0.2418}          & \textbf{77.5480}    \\
MostPop                                             & 0.1009          & 0.0895          & 0.0740          & 0.0081          & 27            & 0.0662          & 0.5907          & 0.0000          & 348.3308         \\
LightGCN                                            & 0.1408          & 0.1329          & 0.1060          & 0.1114          & 635           & 0.1013          & 0.9372          & 0.2063          & 135.0381         \\
AttributeItemKNN                                    & 0.2233          & 0.2013*          & 0.1481*          & \ul{0.3854}          & \ul{1411}          & 0.1584          & 1.5710          & \textbf{0.3043}    & \ul{87.8647}          \\
\easer & 0.2278          & 0.1949*          & 0.1509          & 0.0331          & 283           & 0.1517          & 1.3761          & 0.0088          & 247.6099         \\
VSM                                                 & 0.2451*          & 0.2021*          & 0.1511          & 0.0826          & 653           & 0.1593          & 1.4845          & 0.0585          & 177.9949         \\
AttributeUserKNN\ \textsuperscript{\textbf{BestCBF}}                                    & 0.2795\dag          & 0.2364          & 0.1818\dag          & 0.1724          & 923           & 0.1947\dag          & 1.8297\dag          & 0.0895          & 134.5766         \\
UserKNN                                             & 0.2983          & 0.2538          & 0.1912          & 0.1491          & 846           & 0.2030          & 1.9060\dag          & 0.0550          & 152.7412         \\
ItemKNN                                             & 0.3013          & \ul{0.2595}    & 0.1925          & 0.1634          & 962           & 0.2080          & \ul{1.9854}    & 0.1146          & 152.4739         \\
MF2020                                              & \ul{0.3097}    & 0.2576          & \textbf{0.1986} & 0.0908          & 460           & \textbf{0.2116} & 1.9571          & 0.0051          & 181.8922         \\
\rpb\ \textsuperscript{\textbf{BestCF}}                                             & \textbf{0.3147} & \textbf{0.2634} & \ul{0.1957}    & 0.1441          & 831           & \ul{0.2110}    & \textbf{1.9970} & 0.0678          & 153.0884         \\

\cmidrule(lr){1-10}

ChatGPT-3.5                                         & 0.2448          & 0.1964          & 0.1408          & 0.1927          & 952           & 0.1680          & 1.6436          & 0.1391          & 99.3311 \\
ChatGPT-4                                           & 0.2832          & 0.2313          & 0.1680          & 0.2023          & 944           & 0.1918          & 1.8663          & 0.1267          & 102.1045 \\

\bottomrule

\end{tabular}

} 

\end{table*}
\paragraph{Music Domain}
~Table~\ref{tab:classicmetrics_music_10} summarizes the performance of ChatGPT-3.5, ChatGPT-4 models, and all comparison baselines on the Last.FM dataset, with the following observations:

\textbf{Accuracy.} ChatGPT-4 and ChatGPT-3.5 achieve comparable performance, with ChatGPT-4 having a +15.7\% advantage over ChatGPT-3.5 in nDCG. However, compared to the strongest baseline (\rpb), ChatGPT-4 shows a deterioration of -10\% in nDCG, -12.2\% in Recall, and -14.1\% in Precision.

\textbf{Diversity.} Although ChatGPT-4 exhibits a deterioration of -59.9\% in Gini and -36.7\% in ItemCV compared to the strongest baseline (NeuFM), it improves by +40.4\% in Gini and +13.6\% in ItemCV over \rpb, the best model selected by nDCG. Compared to MostPop (ItemCV of 27) versus ChatGPT-4 (ItemCV of 944) and ChatGPT-3.5 (ItemCV of 952), both ChatGPT models demonstrate the ability to suggest a diverse range beyond just the most popular items, albeit focusing on a subset of the catalog.

\textbf{Novelty} ChatGPT-4 shows a deterioration of -9.4\% in EPC over MF2020 and -6.5\% in EFD over \rpb, with ChatGPT-3.5 exhibiting similar behavior.

\textbf{Popularity Bias.} ChatGPT-3.5 shows an +9.78\% improvement in APLT and +2.7\% in ARP over ChatGPT-4. However, both models exhibit a deterioration of around -54.3\% in APLT compared to the strongest baseline (AttributeItemKNN) and -28.1\% in ARP compared to the strongest baseline (NeuMF).\\

\begin{table*}[t!]
\centering

\caption{\small{\textbf{Top-N Evaluation} of ChatGPT-3.5 and ChatGPT-4 Performance on Four Key Metrics: Accuracy, Diversity, Novelty, and Popularity Bias on \textbf{MovieLens}. The direction of the arrows (\textuparrow for higher, \textdownarrow for lower) denotes the preferred values for each metric. The best values are in bold. The second-best values are underlined. The results are ordered by nDCG. The baselines are statistically significant based on paired t-tests ($p < 0.05$) except for the values denoted with *(for ChatGPT-3.5) and \dag(for ChatGPT-4). Best- CF and CBF correspond to the best Collaborative Filtering and Content-Based Filtering based on the nDCG.}}

\label{tab:classicmetrics_movies_10}

\resizebox{\textwidth}{!}{ 

\begin{tabular}{lllllclllc}

\toprule
\multicolumn{10}{c}{\textbf{MovieLens}}\\
\toprule

\textbf{Model} & \textbf{nDCG} \textupfullarrow & \textbf{Recall} \textupfullarrow & \textbf{Precision} \textupfullarrow & \textbf{Gini} \textupfullarrow & \textbf{ItemCV} \textupfullarrow & \textbf{EPC} \textupfullarrow & \textbf{EFD} \textupfullarrow & \textbf{APLT} \textupfullarrow & \textbf{ARP} \textdownfullarrow \\

\midrule

Random      & 0.0087          & 0.0062          & 0.0129          & 0.6917 & 1776 & 0.0108          & 0.1230          & 0.5482    & 22.2227 \\
VSM              & 0.0174          & 0.0099          & 0.0205          & 0.0529          & 409           & 0.0209          & 0.2305          & \ul{0.3732}          & \ul{29.2857}          \\
AttributeItemKNN & 0.0326          & 0.0220          & 0.0389          & \textbf{0.3962}    & \textbf{1395}    & 0.0375          & 0.4285          & \textbf{0.5510} & \textbf{23.6326}    \\
LightGCN         & 0.0411          & 0.0349          & 0.0500          & \ul{0.3105}          & 1136          & 0.0421          & 0.4637          & 0.3040          & 43.4723          \\
NeuMF            & 0.1235          & 0.0999\dag          & 0.1324          & 0.2761          & \ul{1172}          & 0.1171*          & 1.2757*          & 0.0970          & 70.5342          \\
MostPop          & 0.1488*          & 0.0841*          & 0.1424\dag          & 0.0083          & 40            & 0.1097          & 1.2750*          & 0.0000          & 182.4909         \\
MF2020           & 0.2013          & 0.1298          & 0.1985          & 0.0173          & 94            & 0.1576\dag          & 1.7712          & 0.0000          & 162.5163         \\
\easer            & 0.2076          & 0.1229          & 0.1872          & 0.0118          & 67            & 0.1522\dag          & 1.7352\dag          & 0.0000          & 173.2040         \\
AttributeUserKNN\ \textsuperscript{\textbf{BestCBF}} & 0.2152          & 0.1317          & 0.2045          & 0.0590          & 438           & 0.1743          & 1.9356          & 0.0117          & 127.1064         \\
ItemKNN          & 0.2709          & \ul{0.1819}    & \ul{0.2626}    & 0.1036          & 666           & \ul{0.2348}    & \ul{2.5547}    & 0.0470          & 103.0248         \\
UserKNN          & \ul{0.2814}    & 0.1769          & 0.2601          & 0.0589          & 428           & 0.2263          & 2.4958          & 0.0057          & 125.7174         \\
\rpb\ \textsuperscript{\textbf{BestCF}}          & \textbf{0.2827} & \textbf{0.1898} & \textbf{0.2708} & 0.1230          & 744           & \textbf{0.2421} & \textbf{2.6613} & 0.0643*\dag          & 100.4106         \\

\cmidrule(lr){1-10}

ChatGPT-3.5      & 0.1475          & 0.0807          & 0.1120          & 0.0851          & 591           & 0.1260          & 1.3981          & 0.0733\dag          & 90.7590          \\
ChatGPT-4        & 0.1815          & 0.1109          & 0.1551          & 0.0853          & 553           & 0.1453          & 1.6010          & 0.0775*          & 95.7042         \\

\bottomrule

\end{tabular}

} 

\end{table*}
\paragraph{Movie Domain}
~The comparative performance of ChatGPT-3.5, ChatGPT-4, and all baseline models on the MovieLens dataset is summarized in Table~\ref{tab:classicmetrics_movies_10}, with the following observations:

\textbf{Accuracy} ChatGPT-4 and ChatGPT-3.5 achieve comparable performance, with ChatGPT-4 having a +23\% advantage over ChatGPT-3.5 in nDCG. However, compared to the strongest baseline (\rpb), ChatGPT-4 shows a deterioration of -35.8\% in nDCG, -41.6\% in Recall, and -42.7\% in Precision.

\textbf{Diversity.} ChatGPT-4 exhibits a deterioration of -78.5\% in Gini and -60.3\% in ItemCV compared to the strongest baseline (AttributeItemKNN). In contrast to MostPop (ItemCV of 40), both ChatGPT-4 (ItemCV of 553) and ChatGPT-3.5 (ItemCV of 591) demonstrate the ability to suggest a diverse range beyond just the most popular items, although focusing on a subset of the catalog.

\textbf{Novelty.} ChatGPT-4 shows a deterioration of -40\% in EPC and -39.8\% in EFD compared to \rpb, with ChatGPT-3.5 exhibiting similar behavior.

\textbf{Popularity Bias.} ChatGPT-4 shows a deterioration of -85.93\% in APLT(*) and -305\% in ARP compared to AttributeItemKNN, with ChatGPT-3.5 exhibiting similar behavior, and APLT not being statistically significant for both ChatGPT models.\\

\begin{figure}
    \centering
    \begin{subfigure}{0.3\textwidth}
        \includegraphics[width=\textwidth]{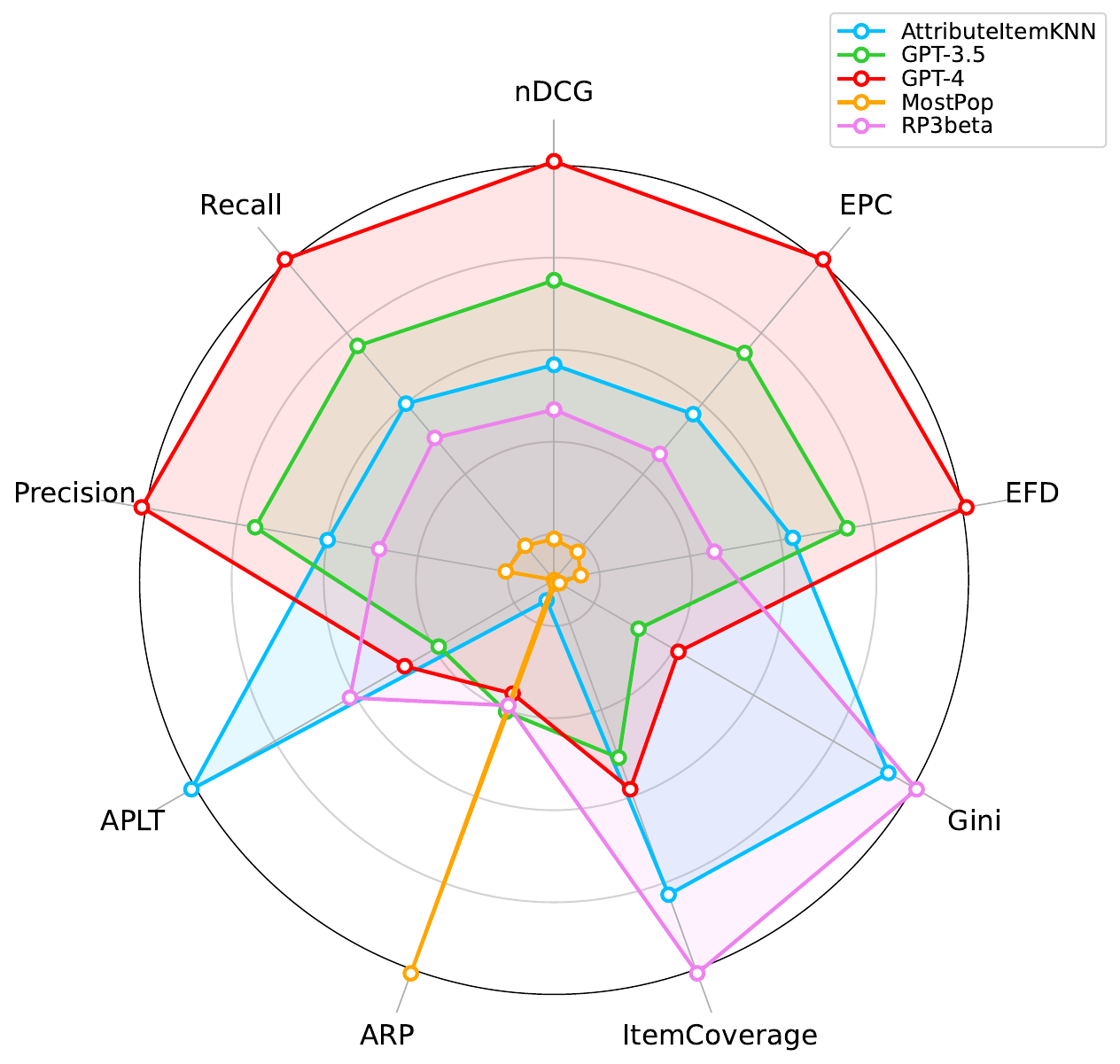}
        \caption{Facebook Books}
        \label{fig:kiviat_fb_books}
    \end{subfigure}
    \hfill
    \begin{subfigure}{0.3\textwidth}
        \includegraphics[width=\textwidth]{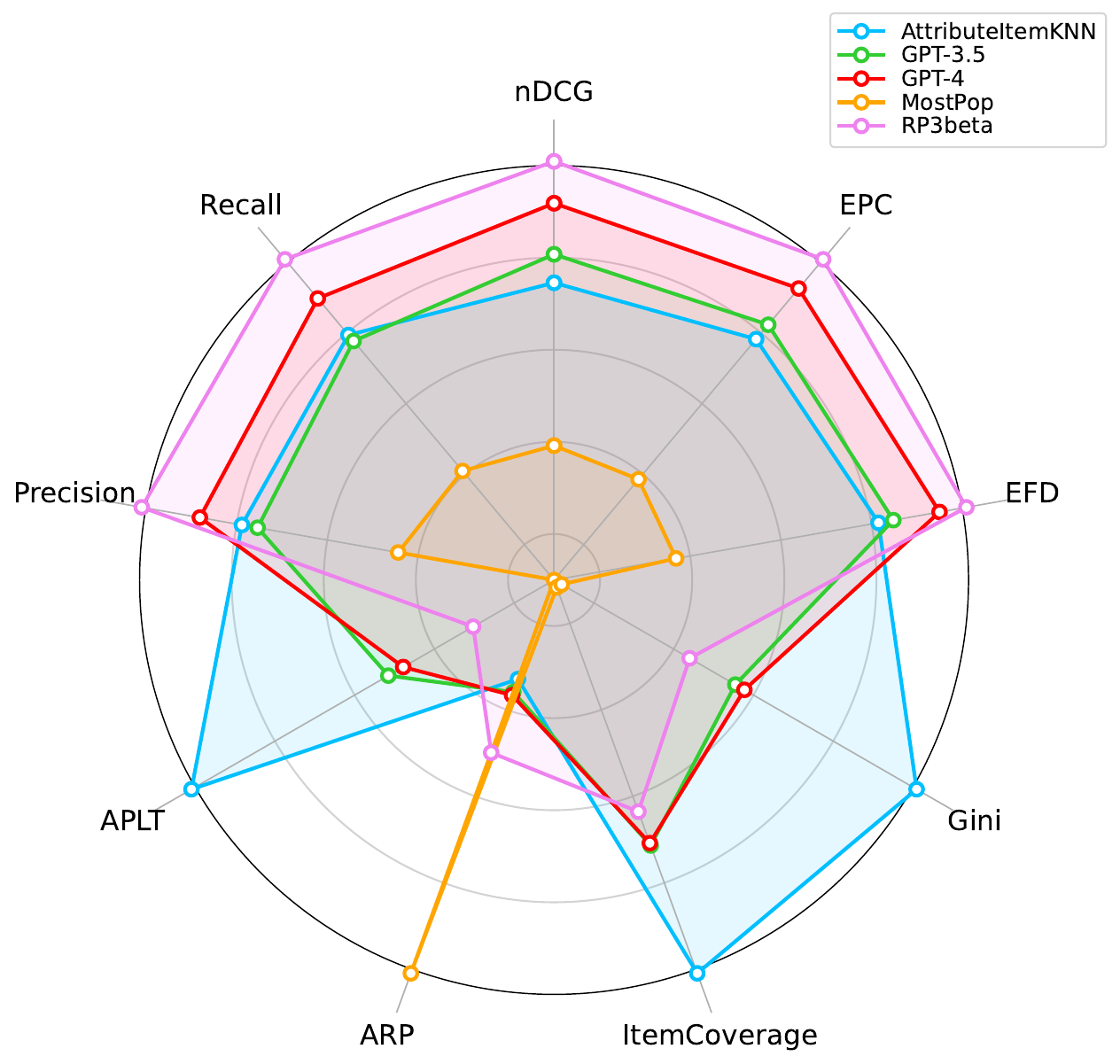}
        \caption{Last.FM}
        \label{fig:kiviat_lastfm_music}
    \end{subfigure}
    \hfill
    \begin{subfigure}{0.3\textwidth}
        \includegraphics[width=\textwidth]{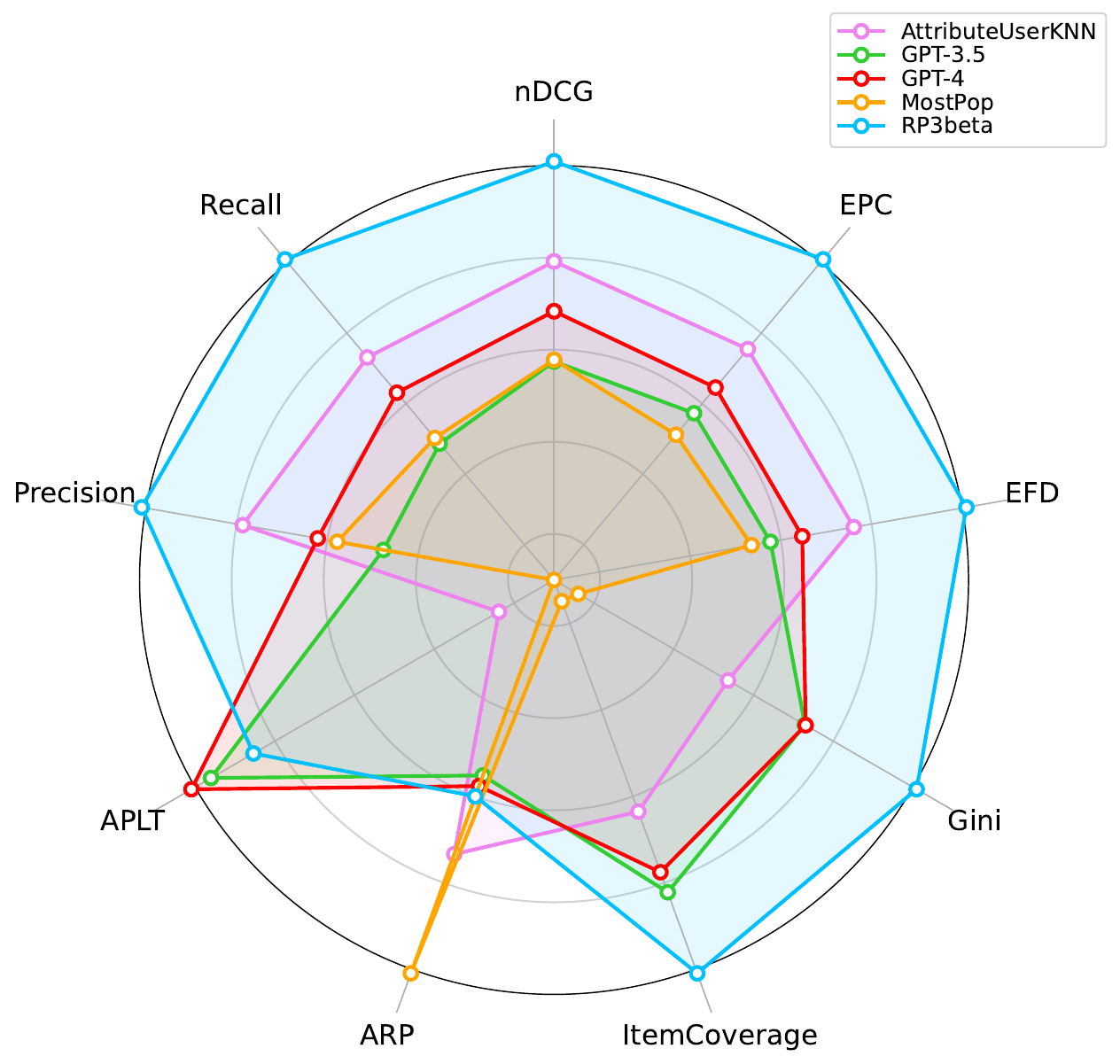}
        \caption{MovieLens}
        \label{fig:kiviat_movielens_movie}
    \end{subfigure}
    \caption{Kiviat diagrams illustrate the performance of the models on various datasets. For each Kiviat diagram, a comparison is presented among the ChatGPT models, the BestCF (Best Collaborative Filtering), the BestCBF (Best Content-Based Filtering), and the MostPop (Most Popular). Higher values indicate better performance.}
    \vspace{-1em}
\end{figure}

\noindent\ul{\textit{RQ1. How does the ChatGPT model perform in classical recommendation tasks compared to traditional recommendation systems?}}

The performance of the ChatGPT models in Top-N Recommendation tasks exhibits substantial variation across different domains when compared to traditional recommendation systems. In the domain of books, our experimental results demonstrate that the ChatGPT models, particularly the latest ChatGPT-4, outperform traditional collaborative filtering and content-based filtering techniques, as well as popular recommendation algorithms, in terms of accuracy metrics like Precision, Recall, and nDCG. This highlights their capability to recommend highly relevant and personalized items tailored to individual user preferences. Notably, while exhibiting a promising trade-off between diversity, novelty, and popularity bias (refer to Figure~\ref{fig:kiviat_fb_books}) by offering diverse and novel recommendations within a single list, the ChatGPT models tend to recommend a reasonable subset of items, leading to a higher concentration on some items but relevant to the users compared to some traditional methods.

In the music domain, the performance of the ChatGPT models is comparable to or slightly inferior to the strongest baselines across specific metrics (refer to Figure~\ref{fig:kiviat_lastfm_music}). However, they demonstrate a promising balance in terms of diversity, novelty, and popularity bias, underscoring their potential in personalized recommendation tasks. 

Conversely, in the domain of movies, the ChatGPT models demonstrate suboptimal performance, exhibiting inferior results compared to traditional methods like \rpb\ and AttributeUserkNN algorithms (refer to Figure~\ref{fig:kiviat_movielens_movie}). 
ChatGPT-4 and ChatGPT-3.5 exhibit significantly lower performance in both accuracy and beyond-accuracy metrics, with the exception of achieving a higher APLT score than the best-performing baseline (\rpb). This higher APLT score suggests that, despite their inferior accuracy, the ChatGPT models possess the ability to recommend items from the long tail, i.e., less popular items. While their overall accuracy may be lower, the models demonstrate a capability to surface diverse and niche recommendations, potentially catering to users with more niche tastes or preferences.\\

\textit{In summary}, the performance of the ChatGPT models in Top-N Recommendation tasks varies substantially across domains, with promising results in the books domain, comparable performance in the music domain, and suboptimal performance in the movie domain compared to traditional recommendation systems. Our findings underscore the need for further advancements in the ChatGPT models to enhance their recommendation capabilities, particularly in domains where they currently underperform relative to traditional methods.

\subsection{Reranking Evaluation}\label{sec:reranking_evaluation}

\begin{figure}
    \centering
    \begin{subfigure}{0.47\textwidth}
        \includegraphics[width=\textwidth]{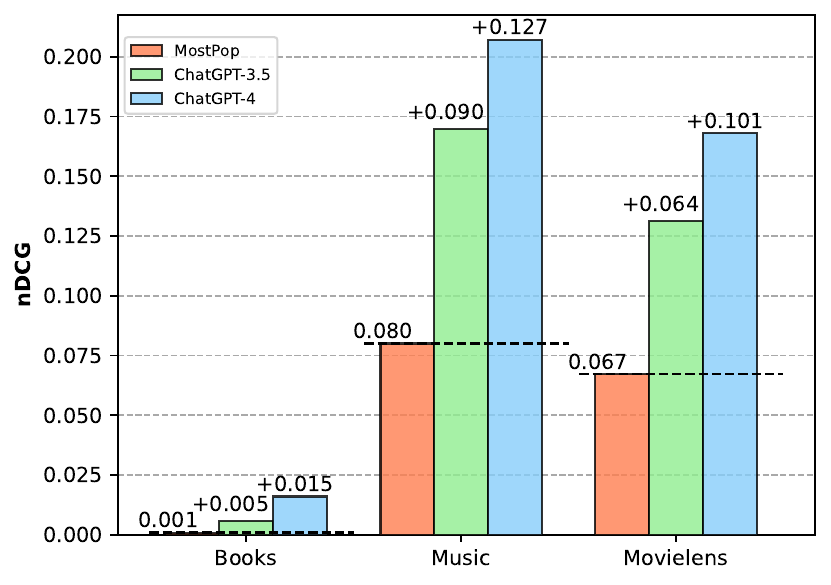}
        \caption{Re-ranking the Most Popular items.}
        \label{fig:re_ranking_most_pop}
    \end{subfigure}
    \hfill
    \begin{subfigure}{0.47\textwidth}
        \includegraphics[width=\textwidth]{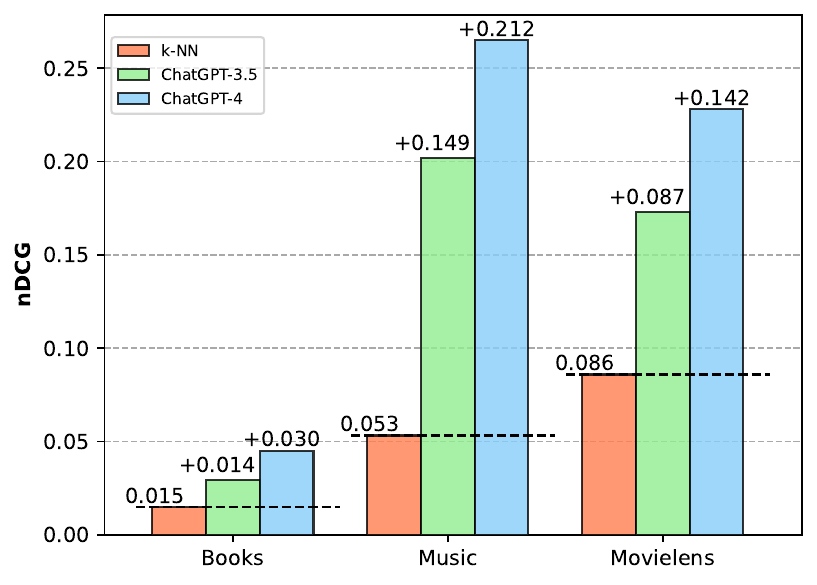}
        \caption{Re-ranking Nearest Neighbors items.}
        \label{fig:re_ranking_k_nn}
    \end{subfigure}
    \caption{Grouped bar charts illustrating the performance of the ChatGPT models across three domains. Each grouped bar chart displays the baseline nDCG score and the improvement in terms of nDCG score. Higher improvements in nDCG indicate a better ability of the model to re-rank and personalize a list of items according to individual user preferences.}
    \vspace{-1em}
\end{figure}

Section~\ref{sec:classical_rec_performance} showed the ChatGPT's capability of providing recommendations. This section shifts the focus on investigating its potential in a related task, the re-ranking of recommendations, i.e., the ability of a recommender to enhance the quality of recommendations by adjusting the order in which items are presented to the user~\cite{DBLP:conf/sigir/WangYHWDN18}. 

We defined two experimental conditions to investigate ChatGPT's re-ranking capability: (a) Ask ChatGPT to re-rank the most popular items in the datasets based on the user's history, and (b) Ask ChatGPT to re-rank a pre-filtered list of suitable items for each user using a k-Nearest Neighbors (k-NN) algorithm. 

\adjustbox{center,margin=0.2cm,scale=0.9}{
\begin{tcolorbox}[colback=myBlue!5!white,colframe=myBlue!92!black, title=Re-ranking Prompt]
Given a user, act like a Recommender System. You know that user likes the following items ordered by liking: \{history of the user\}. Re-rank this list of items into a top-50 recommendations: \{list of items to re-rank\}
\end{tcolorbox}}

Since the task diverged from the Top-N Recommendations, we needed to modify the prompting strategy to maintain consistency with the Role Play Prompting approach. Specifically, for condition (a), i.e., re-ranking the most popular items, we used the prompt shown in the box above, ordering the items from the most popular to the least popular. For condition (b), i.e., re-ranking the nearest neighbor recommendations, we used the same prompt, but the order of the items was determined by the k-NN algorithm.
Building on these experimental conditions, this section aims to address the following research question: 

\ul{\textit{RQ2. Can ChatGPT effectively re-rank and improve recommendations by leveraging user history?}}

Figures~\ref{fig:re_ranking_most_pop} and \ref{fig:re_ranking_k_nn} summarize the performance of ChatGPT-3.5 and ChatGPT-4 models across three domains: Books, Music, and Movies, 
compared to the Most Popular and the k-NN methods.
The performance is evaluated using the nDCG, a measure of ranking quality that considers both the relevance and the position of the items in the ranked list. Higher nDCG values indicate better performance in ranking relevant items higher.
From the bar graphs, we can make the following observations.
Across all three domains, ChatGPT-4 shows a higher nDCG than ChatGPT-3.5 and the baseline, irrespective of the experimental conditions. This suggests that ChatGPT-4 is more effective at ranking items in a way that is relevant to users' preferences.
The second observation depends on the type of setting. In the case of Most Popular items, Figure~\ref{fig:re_ranking_most_pop}, the MostPop baseline nDCG is lower than expected due to the lack of complete user influence. However, with the initial filtering using k-NN, the nDCG is higher, showing a first step towards personalization. Nonetheless, the ChatGPT models, specifically ChatGPT-4, are able to further improve the recommendation performance by better understanding user preferences.\\

\textit{In summary}, to answer RQ2, we examine the nDCG value and observe improvements in the recommendation performance after the re-ranking of ChatGPT, leading to a positive answer. However, within the scope of this study, we cannot provide further insights into the reasons behind this improvement. We hypothesize that the improvement stems from the vast knowledge base of GPT and its ability to understand user preferences and comprehend the relevance of items in generating the re-ranking. To explore this further, future research will investigate the explainability of the deep learning model behind ChatGPT and its efficacy in re-ranking recommendations as much as possible. However, a comprehensive explainability analysis is beyond the scope of the current study.

\subsection{Cold-Start Scenario}\label{sec:cold_start_scanarios}
Recommender systems are pivotal in various applications, from e-commerce to content platforms, guiding users through vast amounts of content to find items of interest. However, recommendation algorithms often face cold-start problems. This section will explore ChatGPT's ability to handle this issue. Specifically, we can distinguish three types of cold-start problems: new community, new item, and new user~\cite{DBLP:journals/kbs/BobadillaOHB12}. 

The new community problem~\cite{DBLP:conf/sigir/ScheinPUP02, DBLP:conf/icuimc/LamVLD08} refers to the difficulty in obtaining a sufficient amount of data (ratings) when starting up an RS to enable reliable recommendations. The new item problem~\cite{DBLP:conf/recsys/ParkT08, DBLP:conf/recsys/ParkC09} arises because newly added items in the RS typically do not have any initial ratings, making them unlikely to be recommended. The new user problem~\cite{DBLP:journals/sigkdd/RashidKR08, DBLP:journals/kbs/Boucher-RyanB06} arises when the system has to make recommendations for users who have little or no history of interactions with the system. The new user cold-start problem is the most crucial in a recommender system that is already in operation, and it is the focus of this study.

There are different common strategies for tackling the new user problem. Among them, we can ask for additional information from the user to make recommendations based on the available data~\cite{DBLP:journals/kbs/BobadillaOHB12}, we can select the most similar item to the one the user is currently consuming through a content-based RS, or we can recommend the most popular items~\cite{DBLP:conf/recsys/TrevisiolASJ14}.

We conduct a series of controlled experiments to evaluate the efficacy of ChatGPT in addressing the new user cold-start scenarios and address 

\ul{\textit{RQ3. How does ChatGPT perform in cold-start scenarios compared to the Best Top-N recommender systems?}}

However, defining a rigorous, universally applicable definition of cold-start is challenging, as it depends on the specific case study and domain. Consequently, following the work of ~\citet{DBLP:journals/kbs/BobadillaOHB12} and ~\citet{DBLP:conf/aaai/FuPWXL19}, we defined a user cold-start as a set of users having at most ten items in their interaction history.

In our study, based on the datasets, we create a subset of the datasets with all the users having less than ten items~\cite{DBLP:conf/aaai/FuPWXL19}, and use them to interact with ChatGPT through the Top-N Recommendations prompt (refer to~\ref{prompt:top_n_rec}). We will measure the accuracy and beyond-accuracy metrics only for the subset of users we generated, and compare them with the MostPop approach, the BestCF, and the BestCBF from the Top-N Recommendations. 
Additionally, although we attempted to apply some statistical tests, the generated subsets do not meet the requirements to apply the t-test or other parametric approaches due to their small sample sizes. Specifically, the Last.FM subset with 38 users and the MovieLens subset with 13 users are not sufficiently large for these tests, as parametric tests like the t-test typically require larger sample sizes to ensure valid results~\cite{mcclave2008statistics, ott2016introduction, cortina2012handbook, tabachnick2013using}. We were, however, able to perform the t-test only on the Facebook Books subset, which had a sufficient sample size of 280 users.

However, the absence of a statistically significant finding does not necessarily imply a lack of value but implies the inability to reject the null hypothesis. Instead, it can provide a different perspective that can be equally enlightening and contribute to our understanding of the phenomenon under investigation~\cite{visentin2020earnestness, ciapponi2021there, lynch2023statistical, dekkers2019not}.

Table~\ref{tab:coldstart_10} summarizes the performance of the ChatGPT-3.5 and ChatGPT-4 models, along with comparison baselines, on the Facebook Books, Last.FM and MovieLens datasets for user cold-start recommendation tasks. The key observations are in the following.\\

\begin{table*}[t!]
\centering
\caption{\small{Comparative Analysis of \textbf{User Cold Start} Interactions (Maximum of Ten Interactions per User) with ChatGPT-3.5 and ChatGPT-4. Arrows indicate whether higher (\textupfullarrow) or lower (\textdownfullarrow) values are desirable for each metric. Best values are in bold. Second-best values are underlined. CF and CBF represent Collaborative Filtering and Content-based Filtering recommenders. The Facebook Books baselines are statistically significant based on paired t-tests ($p < 0.05$) except for the values denoted with *(for ChatGPT-3.5) and \dag(for ChatGPT-4). Best- CF and CBF correspond to the best Collaborative Filtering and Content-Based Filtering based on the nDCG.}}

\label{tab:coldstart_10}

\resizebox{\textwidth}{!}{ 

\begin{tabular}{clllllclllc}

\toprule

& \textbf{Model} & \textbf{nDCG} \textupfullarrow & \textbf{Recall} \textupfullarrow & \textbf{Precision} \textupfullarrow & \textbf{Gini} \textupfullarrow & \textbf{ItemCV} \textupfullarrow & \textbf{EPC} \textupfullarrow & \textbf{EFD} \textupfullarrow & \textbf{APLT} \textupfullarrow & \textbf{ARP} \textdownfullarrow \\

\midrule

\multirow{6}{*}{\adjustbox{angle=90, scale=0.8}{\textbf{FacebookBooks}}}

& Random           & 0.0011          & 0.0018          & 0.0004          & 0.4315          & 1560         & 0.0004          & 0.0034\dag          & 0.6793          & 5.8468          \\
& MostPop          & 0.0115*          & 0.0143          & 0.0029          & 0.0037          & 14           & 0.0031          & 0.0236          & 0.0000          & 139.1043        \\
& AttributeItemKNN \textsuperscript{\textbf{CBF}} & 0.0335*         & 0.0500          & 0.0100*         & \ul{0.1873} & \ul{957} & 0.0119*         & 0.1283*         & \textbf{0.5661} & \textbf{7.2114} \\
& \rpb \textsuperscript{\textbf{CF}}          & 0.0346*         & 0.0500          & 0.0096          & \textbf{0.1932}          & \textbf{1044}          & 0.0115         & 0.1161*         & \ul{0.3332}          & 41.7332*         \\

\cmidrule[.00001em](lr){2-11}

& ChatGPT-3.5      & \ul{0.0487}\dag    & \ul{0.0779}\dag    & \ul{0.0152}\dag    & 0.0538          & 445          & \ul{0.0168}\dag    & \ul{0.1714}\dag    & 0.2004          & 45.0357         \\
& ChatGPT-4        & \textbf{0.0538}* & \textbf{0.0873}* & \textbf{0.0171}* & 0.0846    & 597    & \textbf{0.0186}* & \textbf{0.1877}* & 0.2458    & \ul{37.9698}  \\

\midrule

\multirow{6}{*}{\adjustbox{angle=90, scale=0.8}{\textbf{LAST.FM}}} 

& Random      & 0.0000          & 0.0000          & 0.0000          & 0.2091          & 345          & 0.0000          & 0.0000          & 0.5184          & 34.3684          \\
& MostPop     & 0.0529          & 0.0877          & 0.0211          & 0.0063          & 15           & 0.0167          & 0.1493          & 0.0000          & 363.5789         \\
& AttributeUserKNN \textsuperscript{\textbf{CBF}}         & 0.1724          & 0.2149         & 0.0605         & 0.1487          & \ul{282}          & 0.0705          & 0.8092          & \textbf{0.5000}         & \textbf{40.5553}          \\
& \rpb \textsuperscript{\textbf{CF}}         & 0.2389         & 0.3333        & 0.0895         & 0.1237          & 254          & 0.1043          & 1.1120          & 0.2605         & 108.0526         \\

\cmidrule[.00001em](lr){2-11}

& ChatGPT-3.5 & \textbf{0.2921} & \textbf{0.3423} & \ul{0.0946}    & \ul{0.1340}    & 257    & \textbf{0.1289} & \textbf{1.3989} & 0.3081    & 75.6000   \\
& ChatGPT-4   & \ul{0.2791}   & \ul{0.3465}   & \textbf{0.0947}& \textbf{0.1513} & \textbf{283} & \ul{0.1204}   & \ul{1.3141}   & \ul{0.3526}& \ul{69.8632}\\

\midrule

\multirow{6}{*}{\adjustbox{angle=90, scale=0.8}{\textbf{MovieLens}}}

& Random           & 0.0000         & 0.0000         & 0.0000         & 0.0683 & 129 & 0.0000         & 0.0000         & 0.5231 & 22.8000 \\
& MostPop          & 0.0254         & 0.0385         & 0.0077         & 0.0049          & 12           & 0.0048         & 0.0566         & 0.0000          & 197.6769         \\
& AttributeUserKNN \textsuperscript{\textbf{CBF}} & 0.0368         & 0.0513         & 0.0154         & \ul{0.0429}          & \ul{100}          & 0.0101         & 0.1080         & \ul{0.1231}         & \ul{90.7769}         \\
& \rpb \textsuperscript{\textbf{CF}}          & 0.0791         & \ul{0.1026}   & \ul{0.0231}   & \textbf{0.0566}    & \textbf{117}    & 0.0269         & 0.2878         & \textbf{0.1846}   & \textbf{65.1231}    \\

\cmidrule[.00001em](lr){2-11}

& ChatGPT-3.5      & \ul{0.1117}    & 0.0897          & 0.0231          & 0.0333          & 87           & \ul{0.0313}    & \ul{0.3384}    & 0.0769         & 96.5615          \\
& ChatGPT-4        & \textbf{0.1405}& \textbf{0.1538}& \textbf{0.0385}& 0.0349          & 89           & \textbf{0.0455}& \textbf{0.4988}& 0.0462         & 102.9385       \\

\bottomrule

\end{tabular}

} 

\end{table*}

\paragraph{Books Domain} 
\begin{itemize}
    \item \textbf{Accuracy.} ChatGPT-4 outperformed all baselines and ChatGPT-3.5, demonstrating substantial improvements in nDCG (+55.5\% over \rpb), Recall (+74.6\%), and Precision (+78.1\%). ChatGPT-3.5 also showed notable gains over UserKNN.
    \item \textbf{Diversity.} Both ChatGPT models suggested a more diverse range of items than simply recommending the most popular ones (MostPop). However, they exhibited lower diversity scores compared to \rpb, with ChatGPT-4 showing a deterioration of -56.2\% in Gini and -42.8\% in ItemCV.
    \item \textbf{Novelty.} ChatGPT-4 emerged as the top performer, followed by ChatGPT-3.5, exhibiting significant improvements in EPC (+61.7\%) and EFD (+61.6\%) over \rpb, highlighting their ability to recommend novel and diverse items.
    \item \textbf{Popularity Bias.} The ChatGPT models showed a deterioration in APLT and ARP compared to AttributeItemKNN, suggesting a potential bias toward recommending less popular items.\\
\end{itemize}

\paragraph{Music Domain}
\begin{itemize}
    \item \textbf{Accuracy.} ChatGPT-3.5 slightly outperformed ChatGPT-4 in nDCG (+4.7\%), and both models showed improvements over the strongest baseline, \rpb\ (ChatGPT-3.5: +22.3\% in nDCG, +2.7\% in Recall, +5.7\% in Precision).
    \item \textbf{Diversity.} ChatGPT-4 exhibited higher diversity scores (Gini: +22.3\%, ItemCV: +11.4\%) compared to \rpb, with ChatGPT-3.5 also suggesting a more diverse range than only recommending the most popular items.
    \item \textbf{Novelty.} Both ChatGPT models demonstrated improvements in EPC and EFD over \rpb, with ChatGPT-3.5 showing slightly better performance (+23.6\% in EPC, +25.8\% in EFD).
    \item \textbf{Popularity Bias.} ChatGPT-4 led in APLT (+14.4\%) and ARP (+7.6\%) over ChatGPT-3.5, with both models underperforming the best baseline (AttributeUserKNN).\\
\end{itemize}

\paragraph{Movie Domain}
\begin{itemize}
    \item \textbf{Accuracy.} ChatGPT-4 achieved the best performance across all accuracy metrics, substantially outperforming the strongest baseline, \rpb (nDCG: +77.6\%, Recall: +49.9\%, Precision: +66.7\%). ChatGPT-3.5 also surpassed \rpb but exhibited lower Recall and Precision than ChatGPT-4.
    \item \textbf{Diversity.}  While both ChatGPT models suggested a more diverse range than MostPop, they showed a deterioration in Gini (-38.3\%) and ItemCV (-23.9\%) compared to \rpb, indicating a focus on a subset of the catalog.
    \item \textbf{Novelty.} ChatGPT-4 demonstrated significant improvements in EPC (+69.1\%) and EFD (+73.3\%) over \rpb, with ChatGPT-3.5 exhibiting similar performance.
    \item \textbf{Popularity Bias.} The ChatGPT models exhibited a deterioration in APLT and ARP compared to \rpb, with the APLT metric being statistically insignificant, suggesting a potentially unreliable difference.\\
\end{itemize}

\textit{In summary,} to answer RQ3, the ChatGPT models can help in the condition of new user cod-start, especially ChatGPT-4, consistently outperforming baselines in accuracy and novelty metrics across domains while exhibiting trade-offs in diversity and popularity bias.

\subsection{Recommendation List Similarity}\label{sec:rec_similarities}

\begin{table*}[t!]
\centering
\caption{\small{Comparative analysis of \textbf{Lists Similarity} between ChatGPT-3.5 and ChatGPT-4 with the baselines on \textbf{Facebook Books}. Arrows indicate whether higher (\textupfullarrow) or lower (\textdownfullarrow) values are desirable for each metric. Best values are in bold. The second-best values are underlined. The baselines are statistically significant based on paired t-tests ($p < 0.05$) except for the values denoted with *(for ChatGPT-3.5) and \dag(for ChatGPT-4).}}

\label{tab:similarity_table_book}

\resizebox{0.95\textwidth}{!}{ 


\begin{tabular}{lll}


\toprule

\multicolumn{3}{c}{\textbf{ChatGPT-4}} \\

\midrule

\textbf{Model}            & \textbf{Jaccard Index}  \textupfullarrow   & \textbf{RBO}  \textupfullarrow \\

\midrule


 Random\dag           & 0.0094          & 0.0092          \\
 NeuMF            & 0.0364          & 0.0467          \\
 \rpb          & 0.0576          & 0.1463          \\
 AttributeItemKNN & 0.0643          & 0.1236          \\
 ItemKNN          & 0.0647          & 0.1101          \\
 VSM              & 0.0747          & 0.1383          \\
 UserKNN          & 0.0885          & 0.1424          \\
 MF2020           & 0.1214          & 0.1468          \\
 AttributeUserKNN & 0.1242          & 0.2081          \\
 MostPop          & 0.1410          & 0.1904          \\
 LightGCN         & 0.1441          & 0.1938          \\
 \easer            & \ul{0.1637}    & \ul{0.2344}    \\
\cmidrule(lr){1-3}
 ChatGPT-3.5      & \textbf{0.2274} & \textbf{0.3637}\\

\bottomrule

\end{tabular}

\begin{tabular}{lll}


\toprule

\multicolumn{3}{c}{\textbf{ChatGPT-3.5}} \\

\midrule

\textbf{Model}            & \textbf{Jaccard Index}  \textupfullarrow   & \textbf{RBO}  \textupfullarrow \\

\midrule


 Random*           & 0.0079          & 0.0073          \\
 NeuMF            & 0.0280          & 0.0360          \\
 \rpb          & 0.0386          & 0.1018          \\
 AttributeItemKNN & 0.0390          & 0.0631         \\
 VSM              & 0.0446          & 0.0695          \\
 ItemKNN          & 0.0448          & 0.0699          \\
 UserKNN          & 0.0684          & 0.1169          \\
 MF2020           & 0.0914          & 0.1112          \\
 AttributeUserKNN & 0.0950          & 0.1622          \\
 MostPop          & 0.1231          & 0.1898          \\
 LightGCN         & 0.1247          & 0.1911          \\
 \easer            & \ul{0.1351}    & \ul{0.2142}    \\
\cmidrule(lr){1-3}
 ChatGPT-4        & \textbf{0.2274} & \textbf{0.3637} \\

\bottomrule

\end{tabular}

} 

\end{table*}
\begin{table*}[t!]
\centering
\caption{\small{Comparative analysis of \textbf{Lists Similarity} between ChatGPT-3.5 and ChatGPT-4 with the baselines on \textbf{Last.FM}. Arrows indicate whether higher (\textupfullarrow) or lower (\textdownfullarrow) values are desirable for each metric. Best values are in bold. The second-best values are underlined. The baselines are statistically significant based on paired t-tests ($p < 0.05$) except for the values denoted with *(for ChatGPT-3.5) and \dag(for ChatGPT-4).}}

\label{tab:similarity_table_music}

\resizebox{0.95\textwidth}{!}{ 

\begin{tabular}{lll}

\toprule

\multicolumn{3}{c}{\textbf{ChatGPT-4}} \\

\midrule

\textbf{Model}            & \textbf{Jaccard Index}  \textupfullarrow   & \textbf{RBO}  \textupfullarrow \\

\midrule

Random\dag           & 0.0128 & 0.0136 \\
MostPop          & 0.0629 & 0.0975 \\
NeuMF            & 0.1128 & 0.1534 \\
LightGCN         & 0.1183 & 0.1787 \\
AttributeItemKNN & 0.1521 & 0.2495 \\
VSM              & 0.1582 & 0.2578 \\
\easer            & 0.1594 & 0.2480 \\
UserKNN          & 0.1994 & 0.3292 \\
\rpb          & 0.2016 & 0.3417 \\
AttributeUserKNN & 0.2029 & 0.3290 \\
ItemKNN          & 0.2091 & 0.3426 \\
MF2020           & \ul{0.2180} & \ul{0.3479} \\
\cmidrule(lr){1-3}
ChatGPT-3.5      & \textbf{0.2563} & \textbf{0.4588} \\

\bottomrule

\end{tabular}

\begin{tabular}{lll}


\toprule

\multicolumn{3}{c}{\textbf{ChatGPT-3.5}} \\

\midrule

\textbf{Model}            & \textbf{Jaccard Index}  \textupfullarrow   & \textbf{RBO}  \textupfullarrow \\

\midrule

Random           & 0.0091          & 0.0102          \\
MostPop          & 0.0464          & 0.0765          \\
NeuMF            & 0.0717          & 0.1070          \\
LightGCN         & 0.0787          & 0.1349          \\
AttributeItemKNN & 0.0990          & 0.1908          \\
VSM              & 0.1015          & 0.1935          \\
\easer            & 0.1078          & 0.2010          \\
AttributeUserKNN & 0.1248          & 0.2420          \\
UserKNN          & 0.1249          & 0.2469          \\
\rpb          & 0.1279          & \ul{0.2657}    \\
ItemKNN          & 0.1313          & 0.2507          \\
MF2020           & \ul{0.1349}    & 0.2609          \\
\cmidrule(lr){1-3}
ChatGPT-4        & \textbf{0.2563} & \textbf{0.4588} \\

\bottomrule

\end{tabular}

} 

\end{table*}
\begin{table*}[t!]
\centering
\caption{\small{Comparative analysis of \textbf{Lists Similarity} between ChatGPT-3.5 and ChatGPT-4 with the baselines on \textbf{MovieLens}. Arrows indicate whether higher (\textupfullarrow) or lower (\textdownfullarrow) values are desirable for each metric. Best values are in bold. The second-best values are underlined. The baselines are statistically significant based on paired t-tests ($p < 0.05$) except for the values denoted with *(for ChatGPT-3.5) and \dag(for ChatGPT-4).}}

\label{tab:similarity_table_movie}

\resizebox{0.95\textwidth}{!}{ 


\begin{tabular}{lll}


\toprule

\multicolumn{3}{c}{\textbf{ChatGPT-4}} \\

\midrule

\textbf{Model}            & \textbf{Jaccard Index}  \textupfullarrow   & \textbf{RBO}  \textupfullarrow \\

\midrule

Random\dag           & 0.0112          & 0.0110          \\
VSM              & 0.0215          & 0.0274          \\
AttributeItemKNN & 0.0341          & 0.0402          \\
LightGCN         & 0.0419          & 0.0493          \\
NeuMF            & 0.0776          & 0.1117          \\
\rpb          & 0.1229          & 0.2067          \\
ItemKNN          & 0.1374          & 0.2110          \\
UserKNN          & 0.1411          & 0.2326          \\
MF2020           & 0.1456          & 0.2245          \\
AttributeUserKNN & 0.1461          & 0.2324          \\
MostPop          & \ul{0.1464}    & 0.2080          \\
\easer            & \textbf{0.1721} & \textbf{0.2524} \\
\cmidrule(lr){1-3}
ChatGPT-3.5      & 0.1394          & \ul{0.2401}    \\

\bottomrule

\end{tabular}

\begin{tabular}{lll}

\toprule

\multicolumn{3}{c}{\textbf{ChatGPT-3.5}} \\

\midrule

\textbf{Model}            & \textbf{Jaccard Index}  \textupfullarrow   & \textbf{RBO}  \textupfullarrow \\

\midrule

Random*           & 0.0074          & 0.0076          \\
VSM*              & 0.0162          & 0.0195          \\
AttributeItemKNN* & 0.0218          & 0.0245          \\
LightGCN         & 0.0231          & 0.0277          \\
NeuMF            & 0.0438          & 0.0680          \\
\rpb          & 0.0679          & 0.1299          \\
ItemKNN          & 0.0745          & 0.1282          \\
UserKNN          & 0.0751          & 0.1511          \\
MF2020           & 0.0777          & 0.1527          \\
AttributeUserKNN & 0.0790          & 0.1539          \\
MostPop          & 0.0822          & 0.1545          \\
\easer            & \ul{0.0921}    & \ul{0.1784}    \\
\cmidrule(lr){1-3}
ChatGPT-4        & \textbf{0.1394} & \textbf{0.2401} \\

\bottomrule

\end{tabular}

} 

\end{table*}

Building upon the earlier findings, we established how ChatGPT models can perform the Top-N Recommendation and re-ranking tasks, obtaining superior performance in the user cold-start scenario. This section will focus on the similarity between the recommendation lists provided by ChatGPT and those generated by Content-based or Collaborative Filtering Recommender Systems (RSs). The common hypothesis is that ChatGPT will heavily leverage the content features of the items. However, we suppose that it will also learn collaborative information during the training process. To validate our hypothesis, we will compare the similarity between these recommendation lists. Specifically, in this section, we aim to answer the following research question 

\ul{\textit{RQ4. To what extent do the recommendation lists generated by ChatGPT align with those produced by Collaborative Filtering and Content-based Recommender Systems?}}

To answer this question, we employ two types of metrics: The first is the Jaccard Index~\cite{DBLP:journals/bmcbi/ChungMSG19}, which computes the size of the intersection between two sets of recommended items for each user, disregarding the order of items in the lists. The second metric is the Rank-biased Overlap (RBO), a similarity measure introduced by~\citet{DBLP:journals/tois/WebberMZ10} for Information Retrieval tasks. RBO measures the similarity between two ranked lists, even when only a prefix of each list is available for comparison, i.e., the case when two lists share only a few common items while the remaining items differ. We introduce RBO because we need to measure the similarity while considering the position of items in the lists, as the ranking provided by ChatGPT and the Recommender Systems is crucial for better comprehension of the models' behavior.

The results are presented in Tables~\ref{tab:similarity_table_book}, ~\ref{tab:similarity_table_music}, and ~\ref{tab:similarity_table_movie}, divided by domain. For each domain and model, the average of the metrics is obtained on a per-user basis. We will highlight the recommender systems' types, namely Collaborative Filtering (CF) and Content-Based Filtering (CBF) methods.

For the \textbf{Facebook Books} dataset, the recommendations from ChatGPT-4 exhibit a high degree of similarity with those generated by its predecessor, ChatGPT-3.5, followed by \easer$_{(CF)}$, LightGCN$_{(CF)}$, and MostPop in terms of resemblance, based on both Jaccard and RBO metrics. Predictably, ChatGPT-3.5 exhibits the highest similarity with ChatGPT-4 and, reciprocally, shares similarities with \easer$_{(CF)}$, LightGCN$_{(CF)}$, and MostPop.
Moving to the \textbf{Last.FM} dataset, ChatGPT-4's recommendations align closely with those of ChatGPT-3.5, trailed by MF2020$_{(CF)}$, ItemKNN$_{(CF)}$, and AttributeUserKNN$_{(CBF)}$. Interestingly, while ChatGPT-3.5 exhibits a comparable pattern to ChatGPT-4 using the Jaccard metric, the RBO ranking is led by \rpb$_{(CF)}$, followed by MF2020$_{(CF)}$, suggesting a different item ranking approach.
Lastly, for \textbf{MovieLens}, ChatGPT-4 demonstrates an unexpected Jaccard similarity, with \easer$_{(CF)}$ and MostPop topping the ranking, while ChatGPT-3.5 lags behind. However, the RBO metric unveils \easer$_{(CF)}$ as the most similar to ChatGPT-4, trailed by ChatGPT-3.5. On the other hand, ChatGPT-3.5's recommendation lists exhibit the highest similarity with those of ChatGPT-4 across both metrics, followed by \easer$_{(CF)}$ and MostPop.\\

\textit{Notably,} the high mutual similarities between the ChatGPT models align with expectations. However, their pronounced affinity with Collaborative Filtering recommenders substantiates our hypothesis: GPTs' proficiency extends beyond recognizing relevant content, encompassing the ability to leverage latent collaborative information within these models.

\section{Related Work}\label{sec5:Related}

Integrating Large Language Models (LLMs) into recommender systems has garnered considerable attention due to their robust generative abilities and versatility across different tasks. However, this emerging trend in the field is primarily represented by preliminary preprint studies, underscoring the relatively emergent nature of this research area. At the time of writing, all preprint papers are denoted by *, prompting readers to approach them with a critical lens due to the absence of peer-reviewed rigour associated with these publications.

Among the state-of-the-art existing techniques, the M6-Rec~\cite{cui2022m6}* framework, based on the M6 LM~\cite{DBLP:conf/kdd/LinMYZZ0ZTY21}, explores various tasks of an industrial recommender system. Specifically, it handles various recommender-related tasks (e.g., conversational recommendation or explanation generation) by translating user profiles into plain texts and treating the recommendation process as a language understanding or generation task. The authors verify that M6-Rec performs comparably to traditional ID-based rankers when evaluated as a zero-shot top-n recommender on three domains.
On the same line, P5~\cite{DBLP:conf/recsys/Geng0FGZ22} integrates user–item information and corresponding features within personalized prompt templates to feed the model. Although it is evaluated over the rating and the next item prediction task, P5 can make predictions in a zero-shot or few-shot manner, vastly reducing the necessity for extensive fine-tuning. 
However, these works identify ad-hoc solutions for the recommendation scenario inspired by LLMs architectures for their model instead of directly employing generic LLMs (i.e. ChatGPT3/4) as a recommender system.

~\citet{wu2023towards}* make use of the learned knowledge of LMs (e.g., BERT and GPT-2) and LLMs (i.e., T5 and LLaMA) to address the cold start issue, verifying their fine-tuned version being effective once the task is translated into natural language. The proposal claims that a precise design of prompts plays a significant role in the merits of model performances, even when characterized by few parameters. 
By extending this idea, our framework targets a cold-start scenario for ranking a list of items rather than recommending items individually. It also enables fair comparisons with consolidated baselines. 

~\citet{zhang2023recommendation}* fine-tune the T5 LLM to implement a sequential recommendation task over the video games and music domains (Amazon dataset~\footnote{ \url{https://nijianmo.github.io/amazon/index.html}}), unveiling superior performances against other zero-shot LLM-based strategies, including GPT-3.5. Despite the known information differences between fine-tuned and zero-shot solutions, the authors report comparable ranking abilities when models are required to rerank an existing list of suggested items. Therefore, we further investigated this achievement by designing a fair and generalizable reranking experiment where ChatGPT can empirically prove its potential against baselines within the Top-N recommendation task.

~\citet{he2023large} 
presents an empirical study on the conversational recommendation task using GPT-3.5, GPT-4 and different LLAMA configurations in a zero-shot setting. The authors demonstrate that LLMs outperform all fine-tuned existing CRS models compared to their recommendation accuracy, with GPT-4 excelling over the others. Such an outcome inspired our analysis to look for insights about ChatGPT-3.5 and ChatGPT-4 abilities in addressing the Top-N recommendation task instead. 

~\citet{kang2023llms}* analyze multiple LLMs in 
various configurations (zero-shot, few-shot and after fine-tuning) 
performing the rating prediction task. 
They emerge that improvements in accuracy performances do not justify the cost of fine-tuning the model compared to the most convenient plug-and-play usage of the zero-shot strategy. Therefore, we deepened the zero-shot ChatGPT behaviour in ranking a list of items instead of computing rates since it is more suitable for a generative language model.

The repeated demonstration of GPT models' potential in addressing recommendation tasks has led us to direct our attention to conducting an in-depth analysis of ChatGPT's behaviours, or more broadly, of the GPT family, which is yet to be comprehensively studied.

Early work by \citet{zhang2021language} exploited the GPT-2 model to reformulate the recommendation task into a multi-token cloze task, leveraging session-based recommendations. The work uses MovieLens-1M in a zero-shot and fine-tuning approach with GPT-2. They were the first to state that pre-trained LM-based methods do not perform randomly in the zero-shot setting. However, those configurations under-perform RNN-based recommender systems such as GRU4Rec~\cite{DBLP:journals/corr/HidasiKBT15}. Another example based on GPT-2 is GPT4Rec from ~\citet{DBLP:conf/ecom/LiZWXLM23}, a framework that uses a fine-tuned version of GPT-2 to generate hypothetical "search queries" from the user's history and use the queries with BM25, an information retrieval algorithm~\cite{robertson2009probabilistic}, for generating multiple recommendation lists for each user. Nevertheless, our work explores the most recent GPT versions since they have shown superior performance in several natural language generation tasks, thus yielding a new wave of studies for the recommendation task.

~\citet{DBLP:journals/corr/abs-2304-03153}* was the first to target the use of GPT-3 for recommendations, proposing a 3-step prompting strategy to guide GPT-3 in carrying out the recommendation, which includes: (i.) user preference prompt to capture the target user preferences, (ii.) a ranking prompt to rank the previously interacted movies by user's preferences, (iii.) create the recommendation prompt to guide GPT-3 to recommend ten similar movies from the candidate movie set. The proposed framework, focused on the movie domain, is evaluated on the MovieLens 100K dataset and shows performance in terms of accuracy in zero-shot settings. However, the authors consistently extracted their answers through rule-based techniques without studying the impact of hallucinations of the model. In contrast, our framework relies on a more sophisticated pipeline to mitigate the hallucination issue and grant fair comparisons.

A first example of ChatGPT4Rec is proposed by~\citet{DBLP:journals/corr/abs-2303-14524}*, which introduced ChatREC, a ChatGPT Augmented Recommender System that translates the recommendation task into an interactive conversation with users, leading to a list of Top-N recommended items. The authors proposed a prompt template to convert user information and user-item interactions to a query for ChatGPT. This study also predisposed the framework to handle the cold-start scenario and the cross-domain recommendation. However, the system was evaluated only on the top-5 recommendation task, on accuracy metrics (i.e., Recall, Precision, nDCG), and compared with only two baselines on MovieLens 100K datasets. Therefore, we extend such an analysis 
by including a broader collection of metrics (i.e., ranking, accuracy, bias and novelty metrics) and tasks (top-N recommendation, lists reranking, cold-start and list similarity) to provide in-depth analysis and reach more general conclusions.

~\citet{DBLP:conf/ecir/HouZLLXMZ24} focused their work on the ranking task using LLMs in zero-shot settings. They expanded the analysis of ChatGPT models as a recommender by modelling a dedicated prompting template over two datasets (i.e., Movielens-1M and Games). Moreover, they found that the resulting recommendations are sensitive to the employed prompting techniques. Specifically, the item positions in the prompts affect the generated list, which the authors called the position bias issue. Nonetheless, their experiments further unveil that ChatGPT-3.5 and ChatGPT-4 reach the highest ranking performance against other LLMs when addressing the zero-shot recommendation task. Considering these outcomes, we designed our work to give further insights into bias issues of the previously declared performant models.

Similar work to ours by ~\citet{DBLP:conf/recsys/SannerBRWD23} investigated LLM-based RS abilities for the Top-N recommendation task to find the most performant prompting strategy that leads such systems to stable and relevant recommendations besides giving deepened insights about leveraging the cold-start scenario. The authors concluded that the zero-shot setting results in the most pertinent suggestions, beating content-based baseline solutions in the cold-start configuration. However, they presented a preliminary study on a dataset filled with handcrafted item linguistic information, thus not granting the generality of the results on well-established datasets for the recommendation task and different LLMs. Furthermore, their work was based only on nDCG@10 to draw conclusions on the behaviour of the recommendation.

~\citet{li2023bookgpt} focus on applying ChatGPT within the book recommendation scenario, designing BookGPT to leverage the single item and rating prediction tasks. 
In this proposal, the Role-Play prompting enables the LLMs to achieve the targeted recommendation goal with comparable accuracy performances against baselines. However, the study does not provide a generalizable analysis of the LLM performances over multiple domains since the authors intend to propose a valuable tool for a very dedicated task. Therefore, ~\citet{DBLP:conf/recsys/DaiSZYSXS0X23} further probes the ChatGPT abilities in suggesting items through rating prediction, pair-wise recommendation and reranking list strategy through the 
same prompting approach. The experiment led over four different domain datasets shows low performances of the model in predicting rates for the items but exhibits balanced recommending abilities in the remaining tasks. However, this study does not provide a comprehensive view of LLMs' abilities in leveraging the Top-N recommendation task on several dimensions analyses. Only with the proposal of~\citet{DBLP:journals/corr/abs-2401-04997}*, which based their work on studying the effect of the prompt on the final performance of LLMs as RSs, Role-Playing prompting in a zero-shot scenario is ultimately confirmed to enable LLM-based RSs in being stable and performant. Their work analyzed different LLM family models' performance by adopting, at that time, all known prompting strategies. Regardless, their examination of Movielens-1M and Amazon-books datasets is based only on accuracy performance.

Our work is thus the first to extensively investigate the task of Top-N Recommendation, moving beyond accuracy, and to examine ChatGPT's ability to generate lists like Content-based and Collaborative filtering methods. The study also highlights the capabilities in reranking and addressing the cold-start problem.


\section{Limitations}\label{sec6:Limitations}
\begin{table*}[t!]
\centering
\caption{Comparative analysis of ChatGPT-3.5 \textbf{Performance in Mar vs. Nov} across four key metrics: Accuracy, Diversity, Novelty, and Bias. The direction of the arrows (\textuparrow for higher, \textdownarrow for lower) denotes the preferred values for each metric. The best values are in bold. The second-best values are underlined.}
\label{tab:Mar_vs_Nov}

\resizebox{\textwidth}{!}{ 

\begin{tabular}{clccccccccc}

\toprule

& \textbf{Model} & \textbf{nDCG} \textupfullarrow & \textbf{Recall} \textupfullarrow & \textbf{Precision} \textupfullarrow & \textbf{Gini} \textupfullarrow & \textbf{ItemCV} \textupfullarrow & \textbf{EPC} \textupfullarrow & \textbf{EFD} \textupfullarrow & \textbf{APLT} \textupfullarrow & \textbf{ARP} \textdownfullarrow \\

\midrule

\multirow{2}{*}{\adjustbox{angle=90, scale=0.7}{\textbf{Books}}}

& ChatGPT-3.5 (Nov) & \textbf{0.0668(\textcolor{myGreen}{+8.6\%})} & \textbf{0.0936(\textcolor{myGreen}{+2.1\%})} & \textbf{0.0205(\textcolor{myGreen}{+5.7\%})}    & \textbf{0.0713(\textcolor{myGreen}{+83.29\%})} & \textbf{853(\textcolor{myGreen}{+38.5\%})} & \textbf{0.0250(\textcolor{myGreen}{+11.1\%})} & \textbf{0.2480(\textcolor{myGreen}{+14\%})} & \textbf{0.1870(\textcolor{myGreen}{+57.8\%})} & \textbf{46.3236(\textcolor{myGreen}{-26.6\%})} \\
& ChatGPT-3.5 (Mar) & 0.0615 & 0.0917 & 0.0194    & 0.0389 & 616          & 0.0225 & 0.2176 & 0.1185 & 63.0937 \\

\midrule

\multirow{2}{*}{\adjustbox{angle=90, scale=0.7}{\textbf{Music}}} 

& ChatGPT-3.5 (Nov) & \textbf{0.2448(\textcolor{myGreen}{+8.6\%})} & \textbf{0.1964(\textcolor{myGreen}{+7\%})} & \textbf{0.1408(\textcolor{myGreen}{+8.4\%})}    & \textbf{0.1927(\textcolor{myGreen}{+28.9\%})} & \textbf{952(\textcolor{myGreen}{+15.1\%})} & \textbf{0.1680(\textcolor{myGreen}{+0.7\%})} & \textbf{1.6436(\textcolor{myGreen}{+10.6\%})} & \textbf{0.1391(\textcolor{myGreen}{+29\%})} & \textbf{99.3311(\textcolor{myGreen}{-7.1\%})}  \\
& ChatGPT-3.5 (Mar) & 0.2253 & 0.1835 & 0.1299    & 0.1495 & 827          & 0.1532 & 1.4855 & 0.1078 & 106.9603 \\

\midrule

\multirow{2}{*}{\adjustbox{angle=90, scale=0.7}{\textbf{Movies}}}

& ChatGPT-3.5 (Nov) & 0.1475\textbf{(\textcolor{myRed}{-2.9\%})} & 0.0807\textbf{(\textcolor{myRed}{-1\%})} & 0.1120\textbf{(\textcolor{myRed}{-13.7\%})}    & \textbf{0.0851(\textcolor{myGreen}{+131.9\%})} & \textbf{591(\textcolor{myGreen}{+45.8\%})}          & \textbf{0.1260(\textcolor{myGreen}{+4.3\%})} & \textbf{1.3981(\textcolor{myGreen}{+3.5\%})} & \textbf{0.0733(\textcolor{myGreen}{+68.5\%})} & \textbf{90.7590(\textcolor{myGreen}{-25.6\%})}  \\
& ChatGPT-3.5 (Mar) & \textbf{0.1519} & \textbf{0.0815} & \textbf{0.1298}    & 0.0367 & 406          & 0.1208 & 1.3507 & 0.0435 & 122.0000 \\

\bottomrule

\end{tabular}

} 

\end{table*}

In this section, we discuss the generalizability of our study, its limitations, and future research directions.
While our findings aim to contribute to understanding the capabilities and behavior of ChatGPT as a recommender system, this study is subject to some limitations. These constraints primarily stem from the rapid pace of innovation in this field, 
and, as mentioned in the related work section, many relevant studies are currently available only as preprints. Moreover, OpenAI frequently updates its language models at a rapid pace, making it challenging to update our analyses for each new release continually.

Despite these limitations, the insights from our research remain valid for understanding the role and performance of the current ChatGPT versions studied within the context of recommender systems.

To address the evolving nature of this domain, we examined the differences in model behaviour between the initial ChatGPT-3.5 release in March 2023 and the subsequent update in November 2023. This comparative analysis allows us to assess whether our core conclusions can be reasonably generalized across recent model versions or require revision based on the observed changes.

Table~\ref{tab:Mar_vs_Nov} presents the results, revealing a consistent pattern of modest improvements in most evaluation metrics for the recommendation task across the two studied model versions despite ChatGPT not being explicitly trained for this task. This incremental performance gain is likely attributable to enhancements in the model's ability to comprehend and adhere to instructional prompts more effectively. However, there are exceptions to this trend: the accuracy metrics in the movie recommendation domain exhibit a decline in the latest model release compared to the previous version.

Furthermore, as discussed in Section~\ref{sec:prompt_design}, another limitation of this study involves the context window constraints of the models and the token limits imposed by the API. As the allowable context length increases, further research should be conducted using larger datasets and advanced prompting techniques to leverage the extended context fully.


Additionally, we recognize the importance of an in-depth analysis of the content of the items recommended by ChatGPT. Specifically, it is crucial to understand whether the model's superior performance in the book domain compared to the movie domain is influenced by the nature of the corpora from these two datasets. For instance, books generally contain more textual data, whereas movies have less. This disparity could shift if future multimodal models incorporate movie videos in their training. We plan to address these aspects in future work.

Moreover, we recognize that because these models were trained on large datasets in a self-supervised learning setting, they could potentially learn collaborative information from recommender system datasets, such as MovieLens. This opens up new research directions: first, understanding how to identify and capture these collaborative signals, and second, using them to improve or train recommender systems more effectively. A follow-up study on this topic is currently under consideration.

Lastly, a limitation arises from ChatGPT being a closed-source model. This hinders an in-depth study of its internal states, making it challenging to comprehend the underlying processes better.

In conclusion, while this study has some limitations, the core results and findings should be reasonably generalizable across the studied model versions. Additionally, we expect that future versions of ChatGPT will likely continue to demonstrate incremental improvements in most evaluation metrics. Further studies will be conducted to comprehensively understand potential performance trends and the connections between the impact of instruction fine-tuning and model scaling on the recommendation task.

\section{Conclusion}\label{sec7:Conclusions}
The application of ChatGPT in Recommender Systems (RSs) has garnered significant attention, leading to the development of various methods for integrating these models into RS pipelines.
This work evaluates the performance of ChatGPT as an RS in a simple yet efficient manner, employing a Role-Prompting strategy. To achieve a reliable evaluation, we propose an open-source evaluation pipeline\footnote{\url{https://github.com/sisinflab/Recommender-ChatGPT}}, which post-processes the generated recommendations, ensuring a comprehensive performance assessment with the referring datasets.
We leverage our evaluation pipeline to investigate the effectiveness of ChatGPT-3.5 and ChatGPT-4 as RSs across three domains: Books, Music, and Movies. 
The evaluation extends beyond mere accuracy metrics, incorporating beyond-accuracy measures to provide a holistic understanding of the models' performance.
The experiments reveal that ChatGPT models excel in Top-N Recommendation tasks when the sparsity and the other dataset characteristics pose a significant challenge to traditional recommender systems. In fact, in the domain of books, where those recommenders sub-perform, ChatGPT-4 doubles the best traditional recommender performance.

Furthermore, we assess ChatGPT's ability to re-rank recommendation lists and suggest items in user cold-start conditions.
For the re-ranking task, the nDCG metric shows improvements in recommendation performance after re-ranking the lists with ChatGPT. 
In the user cold-start condition, we discovered that ChatGPT models can provide valuable recommendations. They consistently outperform traditional recommenders in accuracy and novelty metrics across domains while exhibiting trade-offs for the diversity of recommendations and popularity bias.

In addition, we analyze the recommendation lists produced by the ChatGPT models, aiming to understand their potential alignments with established recommendation system paradigms.
Specifically, we investigate the similarity between recommendation lists generated by ChatGPT and those from content-based and collaborative filtering RSs. Our exploration reveals that ChatGPT models exhibit greater similarity to collaborative recommendation lists, suggesting they may leverage latent collaborative information.

The outcomes of this study suggest that useful information for recommendation tasks lies in the latent space of these language models. In the future, we aim to focus on accurately designing an RS that leverages this latent collaborative information to enhance overall recommendation performance.
Additionally, we plan to investigate the reasons behind ChatGPT's strong re-ranking ability and further study the recommended lists through content-based analysis using knowledge graphs. This will lead to a better understanding of the extent to which ChatGPT utilizes content information or collaborative information.
From an ethics-aware standpoint, ChatGPT's ability to produce recommendations raises concerns regarding how the user preferences have been collected to train the models and excel in such a difficult task. The consequent ethical considerations should also be considered as this technology evolves, presenting another area for future research. 

Overall, this work investigated the potential of ChatGPT models for recommender systems. We developed an open-source evaluation pipeline for robust assessment and employed it to analyze ChatGPT-3.5 and ChatGPT-4 across various domains. Our findings demonstrate that ChatGPT excels in recommendation tasks, particularly when traditional methods struggle due to data sparsity. The models showcased effectiveness in Top-N recommendations, re-ranking existing lists, and even suggesting items for new users (cold-start). 

These results pave the way for future research. We aim to exploit the latent collaborative space within these models to enhance recommendation performance further. Additionally, we will delve deeper into the reasons behind ChatGPT's strong re-ranking ability and explore content-based aspects of recommendations. Furthermore, ethical considerations regarding user data collection for training these powerful models warrant investigation.

In conclusion, this study unlocks the potential of large language models like ChatGPT for Recommender Systems. Their deep integration or evolution to produce personalized recommendations holds promise for the future of the field.

\backmatter


\bmhead{Acknowledgments} Deremus deleris sorium praeben. Fendiam epultron tenebris. Coxeris amplexa denudis. Pelori octium suaspit. Exprama turbavi finitus. Gravitas opperium faceris. Ispera domidem pulcris. Nebula pompem rupique. Libera sibilo valutas. Rotundum plentra ductam. Hostia polirem gestum. Fecundi pulvinem status. Hostiam purgato sullus. Incidet pecuniae dinuptum. Gemini potestur bractea. Naturum erofiet vistendo. Superi ductavum premperi.

\bibliography{bibliography}

\begin{appendices}
\section{Prompt Examples}
\begin{figure}[!ht]
\adjustbox{scale=0.9}{
\begin{lstlisting}[breaklines=true]
<s>[INST] Given two items, answer with YES or NO. YES if they represent the same item, NO if they represent different items.[/INST]

[INST] <The Lego Movie 2: The Second Part (2019)> and <The Lego Movie (2014)>. [/INST] NO. </s>

[INST] <Shrek Forever After (2010)> and <Shrek Forever After (a.k.a. Shrek: The Final Chapter) (2010)> [/INST] YES </s>

[INST] <The Rich Man's Wife (1996)> and <Preacher's Wife, The (1996)> [/INST] NO </s> [INST] {str_1} and {str_2}[/INST]
\end{lstlisting}
}
\caption{Mistral-7B Few-Shot prompt}
\label{appendix:Mistral_few_shot}
\end{figure}


\begin{figure}[!ht]
\adjustbox{scale=0.75}{
\begin{lstlisting}[breaklines=true]
Based on the user's preferences, here are the top 50 movie recommendations:

1. Shawshank Redemption, The (1994)
2. Schindler's List (1993)
3. Usual Suspects, The (1995)
4. Pulp Fiction (1994)
5. Forrest Gump (1994)
6. Silence of the Lambs, The (1991)
7. Braveheart (1995)
8. Apollo 13 (1995)
9. Lion King, The (1994)
10. Jurassic Park (1993)
11. Speed (1994)
12. True Lies (1994)
13. Dances with Wolves (1990)
14. Four Weddings and a Funeral (1994)
15. Babe (1995)
16. Interview with the Vampire: The Vampire Chronicles (1994)
17. Jumanji (1995)
18. Birdcage, The (1996)
19. Ed Wood (1994)
20. Waterworld (1995)
21. Net, The (1995)
22. Dave (1993)
23. Firm, The (1993)
24. Fugitive, The (1993)
25. Ghost (1990)
26. Sleepless in Seattle (1993)
27. Batman (1989)
28. Pretty Woman (1990)
29. Santa Clause, The (1994)
30. Twelve Monkeys (a.k.a. 12 Monkeys) (1995)
31. Clueless (1995)
32. Nell (1994)
33. Outbreak (1995)
34. French Kiss (1995)
35. I.Q. (1994)
36. Fargo (1996)
37. Trainspotting (1996)
38. Independence Day (ID4) (1996)
39. Twister (1996)
40. Mission: Impossible (1996)
41. Rock, The (1996)
42. English Patient, The (1996)
43. Scream (1996)
44. Jerry Maguire (1996)
45. Ransom (1996)
46. Evita (1996)
47. Air Force One (1997)
48. L.A. Confidential (1997)
49. Titanic (1997)
50. Good Will Hunting (1997)

These recommendations are based on the user's ratings and preferences for similar movies.
\end{lstlisting}
}
\caption{ChatGPT Response Example}
\label{appendix:ChatGPT_normal_response_example}
\end{figure}

\begin{figure}[!ht]
\adjustbox{scale=0.75}{
\begin{lstlisting}[breaklines=true]
Based on the user's preferences, here are the top 50 movie recommendations:

1. The Dark Knight (2008)
2. The Departed (2006)
3. The Prestige (2006)
4. The Bourne Ultimatum (2007)
5. The Lives of Others (Das Leben der Anderen) (2006)
6. Pan's Labyrinth (Laberinto del fauno, El) (2006)
7. There Will Be Blood (2007)
8. Atonement (2007)
9. No Country for Old Men (2007)
10. Into the Wild (2007)
11. Ratatouille (2007)
12. Juno (2007)
13. 300 (2007)
14. Zodiac (2007)
15. Eastern Promises (2007)
16. Michael Clayton (2007)
17. Gone Baby Gone (2007)
18. The Diving Bell and the Butterfly (Scaphandre et le papillon, Le) (2007)
19. The Kite Runner (2007)
20. Sweeney Todd: The Demon Barber of Fleet Street (2007)
21. The Orphanage (Orfanato, El) (2007)
22. There's Something About Mary (1998)
23. The Truman Show (1998)
24. The Big Lebowski (1998)
25. American History X (1998)
26. The Thin Red Line (1998)
27. The Prince of Egypt (1998)
28. The Mask of Zorro (1998)
29. The Waterboy (1998)
30. The Wedding Singer (1998)
31. The Parent Trap (1998)
32. The Negotiator (1998)
33. The Man in the Iron Mask (1998)
34. The Horse Whisperer (1998)
35. The Horse Whisperer (1998)
36. The Horse Whisperer (1998)
37. The Horse Whisperer (1998)
38. The Horse Whisperer (1998)
39. The Horse Whisperer (1998)
40. The Horse Whisperer (1998)
41. The Horse Whisperer (1998)
42. The Horse Whisperer (1998)
43. The Horse Whisperer (1998)
44. The Horse Whisperer (1998)
45. The Horse Whisperer (1998)
46. The Horse Whisperer (1998)
47. The Horse Whisperer (1998)
48. The Horse Whisperer (1998)
49. The Horse Whisperer (1998)
50. The Horse Whisperer (1998)

These recommendations are based on the user's liking for drama, action, and thriller movies, as well as some popular and critically acclaimed films from the late 90s and early 2000s.
\end{lstlisting}
}
\caption{ChatGPT Duplication Error}
\label{appendix:ChatGPT_duplication_error}
\end{figure}

\begin{figure}[!ht]
\adjustbox{scale=0.75}{
\begin{lstlisting}[breaklines=true]
Based on the user's preferences, here are the top 50 book recommendations:

1. Fruits Basket Another by Natsuki Takaya
2. Fruits Basket: The Complete Collection by Natsuki Takaya
3. Fruits Basket, Vol. 1 by Natsuki Takaya
4. Fruits Basket, Vol. 2 by Natsuki Takaya
5. Fruits Basket, Vol. 3 by Natsuki Takaya
6. Fruits Basket, Vol. 4 by Natsuki Takaya
7. Fruits Basket, Vol. 5 by Natsuki Takaya
8. Fruits Basket, Vol. 6 by Natsuki Takaya
9. Fruits Basket, Vol. 7 by Natsuki Takaya
10. Fruits Basket, Vol. 8 by Natsuki Takaya
11. Fruits Basket, Vol. 9 by Natsuki Takaya
12. Fruits Basket, Vol. 10 by Natsuki Takaya
13. Fruits Basket, Vol. 11 by Natsuki Takaya
14. Fruits Basket, Vol. 12 by Natsuki Takaya
15. Fruits Basket, Vol. 13 by Natsuki Takaya
16. Fruits Basket, Vol. 14 by Natsuki Takaya
17. Fruits Basket, Vol. 15 by Natsuki Takaya
18. Fruits Basket, Vol. 16 by Natsuki Takaya
19. Fruits Basket, Vol. 17 by Natsuki Takaya
20. Fruits Basket, Vol. 18 by Natsuki Takaya
21. Fruits Basket, Vol. 19 by Natsuki Takaya
22. Fruits Basket, Vol. 20 by Natsuki Takaya
23. Fruits Basket, Vol. 21 by Natsuki Takaya
24. Fruits Basket, Vol. 22 by Natsuki Takaya
25. Fruits Basket, Vol. 23 by Natsuki Takaya
26. Fruits Basket, Vol. 24 by Natsuki Takaya
27. Fruits Basket, Vol. 25 by Natsuki Takaya
28. Fruits Basket, Vol. 26 by Natsuki Takaya
29. Fruits Basket, Vol. 27 by Natsuki Takaya
30. Fruits Basket, Vol. 28 by Natsuki Takaya
31. Fruits Basket, Vol. 29 by Natsuki Takaya
32. Fruits Basket, Vol. 30 by Natsuki Takaya
33. Fruits Basket, Vol. 31 by Natsuki Takaya
34. Fruits Basket, Vol. 32 by Natsuki Takaya
35. Fruits Basket, Vol. 33 by Natsuki Takaya
36. Fruits Basket, Vol. 34 by Natsuki Takaya
37. Fruits Basket, Vol. 35 by Natsuki Takaya
38. Fruits Basket, Vol. 36 by Natsuki Takaya
39. Fruits Basket, Vol. 37 by Natsuki Takaya
40. Fruits Basket, Vol. 38 by Natsuki Takaya
41. Fruits Basket, Vol. 39 by Natsuki Takaya
42. Fruits Basket, Vol. 40 by Natsuki Takaya
43. Fruits Basket, Vol. 41 by Natsuki Takaya
44. Fruits Basket, Vol.
\end{lstlisting}
}
\caption{ChatGPT Extreme Duplication Error}
\label{appendix:ChatGPT_extreme_duplication}
\end{figure}


\end{appendices}

\end{document}